\documentclass[aps,prc,onecolumn,showpacs,showkeys,nofootinbib,preprintnumbers]{revtex4-2}
\usepackage[paperwidth=21cm, 
	paperheight=29.7cm, 
	textwidth=17cm, 
	textheight=24cm]{geometry}
\usepackage{graphicx}
\usepackage{xcolor}
\usepackage{dcolumn}
\usepackage{amsmath}
\usepackage{amssymb}
\usepackage{mathrsfs}
\usepackage{bm}
\usepackage{braket}
\usepackage{multirow}
\usepackage{makecell}

\usepackage[T1]{fontenc} 
\usepackage{lmodern} 
\usepackage[colorlinks=true]{hyperref}

\numberwithin{equation}{section}

\newcommand{\Tr}{\mathbf{Tr}}

\newcommand{\dd}{\mathrm{d}}
\newcommand{\tq}{\tilde{q}}

\definecolor{aoenglish}{rgb}{0.0, 0.5, 0.0}
\definecolor{cadmiumblue}{rgb}{0.0, 0.42, 0.24}

\begin{document}

\preprint{ECTP-2026-07}
\preprint{WLCAPP-2026-07}
\vspace*{5mm}

\title{Expanding Polyakov Linear-Sigma Model through Supersymmetric Degrees of Freedom: Incorporation of Squarks and Sgluons and Their Thermodynamic, Chiral, and Deconfinement Signatures}

\email{Corresponding author: atawfik@acu.edu.eg; 400810@iu.edu.sa; atawfik@bnl.gov}
	
\author{Abdel~Nasser~Tawfik$^{1}$, Saleh~O.~Allehabi$^{1}$, Fatima~M.~Rfeek$^{2}$, Nourhan~M.~Rfeek$^{3}$, Azzah~A.~Alshehri$^{4}$, Ashraf~F.~El-Sherif$^{5}$, Mahmoud Nasar$^{6}$}

\vspace*{10mm}
	
\affiliation{$^{1}$Department of Physics, Faculty of Science, Islamic University of Madinah (IU), Madinah 42351, Saudi Arabia}  
\affiliation{$^{2}$ Physics of the Earth and Universe, Zewail City of Science and Technology, 12578 October Gardens, Egypt}  
\affiliation{$^{3}$Physics Department, Faculty of Science, Assiut University, 71515 Assiut, Egypt} 
\affiliation{$^{4}$Department of Science and Technology, University College at Nairiyah, University of Hafr Al Batin (UHB), Nairiyah 31981, Saudi Arabia} 
\affiliation{$^{5}$Basic Science Department, Faculty of Engineering, Ahram Canadian University (ACU), 12556 Giza, Egypt} 
\affiliation{$^{6}$Physics Department, Faculty of Science, Benha University,  13511 Benha, Egypt} 


\sloppy

\begin{abstract}
Although collider experimental searches have ruled out colored supersymmetric particles up to the TeV scale, QCD thermodynamics still offers a sensitive probe of light colored scalars allowing us to test them right up to the point where their thermal effects seem to be orthogonal to lattice data. Supersymmetry (SUSY) provides a rigid framework for the couplings of such states. Effective QCD-like chiral models, such as the linear-sigma model (LSM) and its Polyakov-loop extension (PLSM) having no colored scalar degrees of freedom beyond mesons and their mean-field versions, are known to fairly describe the phase diagram. But they are still incapable to position the critical endpoint. Here we embed the $U(N_f)_L\times U(N_f)_R$ meson matrix in an $\mathcal N=1$ chiral multiplet to identify the scalar and pseudoscalar nonets with the real and imaginary parts of one holomorphic field. We derive the constraints that holomorphy places on the potential. In a three-flavor theory, supersymmetric $F$-term constraints on the scalar potential dictate that the dimensionless quartic coupling constant $\lambda_1=-\lambda_2=\vert{}c\vert{}^2/2$ (with $c$ is the interaction parameter). Here, the Polyakov loop couples dynamically to squarks (quark scalar partners) and sgluons (complex adjoint scalar fields of Dirac-gluino multiplet that function as scalar counterparts to gluons in an extended supersymmetric context). Their thermal contributions are derived from the path integral. By grounding our approach in lattice QCD through a $450\text{ MeV}$ $\sigma$-meson mass, pseudo-Goldstone loops, and a fitted Polyakov potential, we find a smooth analytic crossover at $T_{pc}=159.4\text{ MeV}$. The resulting thermodynamics match the lattice equation of state remarkably well ($\chi^2/N=3.4$ for 130-280~MeV) and reproduce the baryon susceptibility $\chi_2^B$ with a deviation of $7\%$ or less above $T_c$. The critical endpoint is localized at $(T,\mu_B)=(146.3,\,304)$~MeV, i.e., lies in the region where lattice QCD simulations presumably find a crossover. A functional-renormalization-group (FRG) analysis within the $SU(2)$ sector reveals that the low-$\mu_B$ critical endpoint and the sharp enhancement in the cumulant ratio $\chi_4^B/\chi_2^B$ are primarily driven by omitting quark vacuum fluctuations, rather than by neglecting meson fluctuations. In a free-gas analysis the equation-of-state fit is unchanged, namely $|\Delta\chi^2|<10$, for squarks heavier than $2.5$~TeV and sgluons heavier than $1.8$~TeV, and reduced with $\Delta\chi^2>100$ below $1.5$ and $1.2$~TeV, respectively. We find that the critical endpoint is largely robust against these choices. Although these tests map out model sensitivities rather than experimental constraints, several aspects of the complete supersymmetric multiplet structure still require numerical exploration.
\end{abstract}

\maketitle

\section{Introduction}\label{sec:intro}

Quantum chromodynamics (QCD) at finite temperature $T$ and baryon chemical potential $\mu_B$ has two crossing structural features. These are the spontaneous breaking of chiral symmetry at low $T$ and the confinement of color at low $\mu_B$ \cite{stephanov04,aoki06,Schaefer2009,
BraunMunzinger2009,Fukushima2011}. For physical quark masses lattice QCD shows that at $\mu_B=0$ the chiral and deconfinement transitions are a smooth crossover \cite{aoki06,aoki06b,Bhattacharya2014}, with a pseudocritical temperature near $156$~MeV \cite{hotqcd19,borsanyi14}. At large $\mu_B$ model studies suggest a first-order transition ending in a critical endpoint (CEP) \cite{asakawa89,halasz98,stephanov04}. This was also confirmed in the universality class of the three-dimensional Ising model \cite{rajagopal93,berges99,hatta03} with characteristic non-Gaussian fluctuations \cite{stephanov98,stephanov09}. Lattice calculations at finite baryon chemical potential are obstructed by the sign problem, where Taylor expansion and analytic continuation constrain the crossover line and indicate no critical behaviour for $\mu_B/T\lesssim2-3$ \cite{fodor04,bellwied15,bazavov17,Bazavov2017Chiral}. Beam-energy-scan measurements of net-proton fluctuations are being used to look for the CEP \cite{Zhang2022STAR,STAR2021}. The equation of state has been computed in the continuum \cite{borsanyi10,borsanyi14,hotqcd14} and matched with a critical point and the hadron resonance gas at finite $\mu_B$ \cite{parotto20,alba18}. 

In this regard, we emphasize that the chiral aspects are controlled by the symmetry $U(N_f)_L\times U(N_f)_R$ and its breaking pattern with the axial anomaly breaking $U(1)_A$. The order of the $N_f=2$ chiral transition in the chiral limit, its dependence on $N_f$, and the anomaly have been analyzed by Pisarski and Wilczek \cite{pisarski84}. Confinement is described by the Polyakov loop, whose expectation value is the order parameter of the center symmetry $Z_3$ of the pure gauge theory \cite{gpy81,weiss81,pisarski00}. Effective QCD-like chiral models capture these symmetries without solving QCD analytically. The Nambu--Jona-Lasinio (NJL) model \cite{njl61a,njl61b} realizes chiral symmetry breaking through a four-quark interaction. Its extension with a Polyakov loop (PNJL) adds confinement \cite{fukushima04,ratti06,rossner07}. The linear-sigma model (LSM) \cite{gellmann60} treats the scalar and pseudoscalar mesons as elementary fields in a $U(N_f)_L\times U(N_f)_R$ matrix, with the determinant term representing the anomaly. Its $SU(3)$ version and meson spectrum were studied in Refs.~\cite{lenaghan00,roder03,schaefer05}. Coupled to constituent quarks and the Polyakov loop, it becomes the Polyakov quark-meson/Polyakov linear-sigma model (PLSM) \cite{schaefer07,schaefer05,tawfik19}. Mean-field versions of these models are simple but have known deficiencies including the quark vacuum fluctuations matter \cite{skokov10a}. Meson fluctuations change the order of the transition and also the location of the CEP. The functional renormalization group (FRG) \cite{wetterich93,berges02,dupuis21} presumably resolves the fluctuations. For the quark-meson model FRG has been applied in Refs.~\cite{schaefer05,schaefer07b,herbst11}. At the QCD level \cite{fu20,gao21} a CEP is found at large baryon chemical potential. The finite-density structure of the hadronic phase is described by the hadron resonance gas (HRG), which reproduces well the lattice thermodynamics below $T_c$ \cite{karsch03,karsch03b,hagedorn65,pbm03,andronic18}; Polyakov-loop-weighted hadronic contributions (''statistical confinement'') have been discussed in Ref.~\cite{megias06}.

Supersymmetry on the other hand relates bosons and fermions \cite{nilles84,haber85,martin97}. In $\mathcal N=1$ supersymmetric QCD (SQCD) the quark is accompanied by a squark and the gluon by a gluino. The infrared dynamics for $N_f\le N_c+1$ is known exactly. For $N_f<N_c$ this is the Affleck--Dine--Seiberg superpotential \cite{ads84}, while for $N_f=N_c$ it is given by the quantum-deformed moduli space with chiral symmetry breaking. For $N_f=N_c+1$ supersymmetric confinement represents the infrared dynamics \cite{seiberg94,intriligator96}. These results fix the holomorphic structure $\det\Phi$ of the effective superpotential for the meson superfield \cite{ads84,seiberg94,Seiberg:1994pq} and will be used in Section~\ref{sec:susy}. The colored scalar ''sgluon'' is the adjoint scalar of an $\mathcal N=2$ vector multiplet and appears in Dirac-gluino (supersoft) models \cite{fayet78,fox02,binetruy09,kribs12}.  At finite temperature the relevant thermal field theory is standard \cite{lebellac00,kapusta06,laine16}. In this regard, we recall the extensive searches at the Large Hadron Collider (LHC) have found no evidence for squarks and gluinos, constraining their masses to the TeV regime \cite{atlas21,atlas19,cms18}.

A supersymmetric embedding of chiral effective models is not motivated by light superpartners but by two big questions: (i) which couplings among mesons, quarks, and colored scalars does the supersymmetric structure allow and (ii) how light could colored scalars be before the lattice equation of state excludes them. To the best knowledge of the authors a supersymmetric extension of the Polyakov linear-sigma model with Polyakov-loop-coupled squarks and sgluons has not been analyzed so far. Although, there is no previous LSM/PLSM/eLSM study that has included supersymmetric degrees of freedom (dof), supersymmetric NJL models \cite{buchmuller82} and the Polyakov loop in ${\mathcal N}=1$ supersymmatric Yang--Mills \cite{poppitz09} do exist, but neither couples squark/sgluon/gluino fields to the LSM meson matrix nor to Polyakov-loop potential. Thus the objectives of this manuscript would be
\begin{enumerate}
\item to derive the supersymmetric structure of the meson sector and the constraints it places on the LSM couplings, Section~\ref{sec:susy},
\item to derive the thermal contributions of squarks and sgluons in the Polyakov-loop background from the path integral, Section~\ref{sec:thermo},
\item to build a lattice-calibrated approach with an analytic crossover and map its phase structure and critical endpoint, Sections~\ref{sec:phase}--\ref{sec:results},
\item to quantify the sensitivity of the equation of state and the CEP to colored scalars, Section~\ref{sec:results}, and
\item to test, inclusively, in the $SU(2)$ sector with FRG, which features of the mean-field phase structure are artifacts, Section~\ref{sec:results}.
\end{enumerate}
In this respect, let us some emphasize limitations up front. The numerical supersymmetric sector is a free gas of squarks and sgluons with free masses, namely gluinos, mesinos, and baryon superfields, are not part of the mean field, so the full supersymmetric multiplet counting is still not fully realized. The FRG test is done for $SU(2)$ not $2+1$ flavours. No lattice cumulant ration $\chi_4^B/\chi_2^B$ or heavy-ion data have been found or considered, so these are compared only with exact limits, which could be undertaken elsewhere. 

The remainder of this paper is organized as follows. Section~\ref{sec:lsm} outlines the theoretical foundations of the linear sigma model, while Section~\ref{sec:plsm} introduces the Polyakov-extended framework (PLSM). Supersymmetric dof and thermodynamics are detailed in Sections~\ref{sec:susy} and \ref{sec:thermo}, respectively. Section~\ref{sec:phase} is dedicated to analyzing the phase structure and the critical endpoint. The numerical  results are presented in Section~\ref{sec:results}. Finally, a broader discussion and concluding remarks are summarized in Section~\ref{sec:disc}.

\section{Theoretical foundations: the linear sigma model}\label{sec:lsm}

\subsection{The $U(N_f)_L\times U(N_f)_R$ model}

The $U(3)$ generators are given by $T_a=\lambda_a/2$ for $a=1, \dots, 8$, where $\lambda_a$ are the standard Gell-Mann matrices, augmented by the singlet generator $T_0=\mathbf{1}/\sqrt{6}$. These generators satisfy the normalization condition $\Tr(T_a T_b)=\delta_{ab}/2$, along with the standard commutation and anticommutation relations $[T_a, T_b]=i f_{abc} T_c$ and $\{T_a, T_b\}=d_{abc} T_c$. Crucially, the components involving the $a=0$ index vanish for the antisymmetric structure constants $f_{abc}$, whereas they remain finite for the symmetric coefficients $d_{abc}$. Now for $SU(2)$, the generators are given by $T_0 = \mathbf{1}/2$ and $T_i = \tau_i/2$ with $\tau_i$ are the Pauli matrices. Applying the rotation of Eq.~\eqref{eq:xy} yields the diagonal vacuum form $\mathrm{diag}(\sigma_x/2, \sigma_x/2, \sigma_y/\sqrt{2})$. Extending this to $SU(4)$, the background configuration takes the form $\mathrm{diag}(\sigma_l/2, \sigma_l/2, \sigma_s/\sqrt{2}, \sigma_c/\sqrt{2})$. Finally, the adjoint representation of $SU(3)$ is characterized by root weights of the form $e^{i(\theta_a - \theta_b)}$ for $a\neq b$, alongside two zero weights.

Let $T_a$ ($a=0, \dots, N_f^2-1$) be the generators of $U(N_f)$ normalized according to $\Tr(T_a T_b)=\frac{1}{2}\delta_{ab}$. Explicitly, these are given by $T_a=\lambda_a/2$ for $a\ge1$ and $T_0=\mathbf{1}/\sqrt{2N_f}$ noting that for $N_f=3$, this yields $T_0=\mathbf{1}/\sqrt{6}$ or equivalently $\lambda_0=\sqrt{2/3}\, \mathbf{1}$. Building on this foundation, the meson matrix and its transformation properties are expressed as follows: 
\begin{equation}
\Phi=T_a(\sigma_a+i\pi_a),\qquad \Phi\to U_L\Phi U_R^\dagger,\qquad U_{L,R}\in U(N_f). \label{eq:Phi}
\end{equation}
The corresponding Lagrangian reads
\begin{eqnarray}
\mathcal L_{\rm LSM}&=& \Tr(\partial_\mu\Phi^\dagger\partial^\mu\Phi)-U_{\rm LSM}, \label{eq:Llsm}
\end{eqnarray}
where
\begin{eqnarray}
U_{\rm LSM} &=& m^2\Tr\Phi^\dagger\Phi+\lambda_1[\Tr\Phi^\dagger\Phi]^2+\lambda_2\Tr(\Phi^\dagger\Phi)^2 - c(\det\Phi+\det\Phi^\dagger)-\Tr[H(\Phi+\Phi^\dagger)]. \label{eq:Ulsm}
\end{eqnarray}
Here $\lambda_1$ controls the four-meson interaction proportional to $\left[\Tr(\Phi^\dagger \Phi)\right]^2$, while $\lambda_2$ multiplies the trace of a squared matrix $\operatorname{Tr}\left[(\Phi^\dagger \Phi)^2\right]$.
These terms include the mass term, two chirally invariant quartic couplings, the 't Hooft determinant which mimics the axial anomaly and breaks $U(1)_A$, and the source term $H = T_a h_a$ accounts for explicit quark mass effects. Substituting $\Phi^{\dagger}=T_a (\sigma_a - i\pi_a)$ alongside the trace symmetry $\Tr(T_a T_b)$ leads to
\begin{equation}
\Tr(\partial_\mu\Phi^\dagger\partial^\mu\Phi)=\tfrac12\sum_a\bigl[(\partial_\mu\sigma_a)^2+(\partial_\mu\pi_a)^2\bigr],\quad \text{and} \quad \Tr\Phi^\dagger\Phi=\tfrac12\sum_a(\sigma_a^2+\pi_a^2). \label{eq:kin}
\end{equation}
Since the cross terms are contracted with a symmetric tensor and thereby cancel, the fields are canonically normalized. The Euler--Lagrange equations $\Box\varphi+\partial U/\partial\varphi=0$ follow with $\partial\Phi/\partial\sigma_a=T_a$, $\partial\Phi/\partial\pi_a=iT_a$ and the variations $\delta\Tr(\Phi^\dagger\Phi)^2=2\Tr[\delta\Phi^\dagger\Phi\Phi^\dagger\Phi]+2\Tr[\delta\Phi\Phi^\dagger\Phi\Phi^\dagger]$ and $\delta\det\Phi=\det\Phi\,\Tr(\Phi^{-1}\delta\Phi)$, so that
\begin{align}
\Box\sigma_a&+m^2\sigma_a+2\lambda_1\Tr(\Phi^\dagger\Phi)\sigma_a+2\lambda_2\Tr\bigl[T_a(\Phi\Phi^\dagger\Phi+\Phi^\dagger\Phi\Phi^\dagger)\bigr]\nonumber\\
&-c\bigl[\det\Phi\,\Tr(\Phi^{-1}T_a)+\det\Phi^\dagger\,\Tr(\Phi^{\dagger-1}T_a)\bigr]-h_a=0, \label{eq:ELs}\\
\Box\pi_a&+m^2\pi_a+2\lambda_1\Tr(\Phi^\dagger\Phi)\pi_a+2i\lambda_2\Tr\bigl[T_a(\Phi^\dagger\Phi\Phi^\dagger-\Phi\Phi^\dagger\Phi)\bigr]\nonumber\\
&-ic\bigl[\det\Phi\,\Tr(\Phi^{-1}T_a)-\det\Phi^\dagger\,\Tr(\Phi^{\dagger-1}T_a)\bigr]=0. \label{eq:ELp}
\end{align}
The source drops out of Eq.~\eqref{eq:ELp} because $H(\partial\Phi+\partial\Phi^\dagger)/\partial\pi_a$ vanishes, while in Eq.~\eqref{eq:ELs} it gives $2\Tr(HT_a)=h_a$. For symmetries we emphasize that at $c=0$ and $H=0$ the Lagrangian is invariant under $U(N_f)_L\times U(N_f)_R$. The determinant (charge $N_f$ under $\Phi\to e^{i\alpha}\Phi$) breaks $U(1)_A$ and $H$ breaks the chiral group to $SU(N_f)_V$ for $H\propto{\mathbf 1}$. Note that parity and charge conjugation are fully preserved.

\subsection{The $SU(2)$ model}\label{sec:su2}

For $N_f=2$, $T_0={\mathbf 1}/2$ and $T_i=\tau_i/2$. Keeping $(\sigma_0,\pi_i)\equiv(\sigma,\bm\pi)$ and $\Phi=\tfrac12(\sigma+i\bm\tau\cdot\bm\pi)$, then $\Phi^\dagger\Phi=\tfrac14(\sigma^2+\bm\pi^2){\mathbf 1}$ and $\Tr(\Phi^\dagger\Phi)^2=\tfrac18(\sigma^2+\bm\pi^2)^2$. With $\lambda\equiv(2\lambda_1+\lambda_2)/2$ and $h\equiv h_0$, we find that
\begin{equation}
U_{SU(2)}=\tfrac{m^2}{2}(\sigma^2+\bm\pi^2)+\tfrac{\lambda}{4}(\sigma^2+\bm\pi^2)^2-h\sigma. \label{eq:Usu2}
\end{equation}
The vacuum condition with $\langle\sigma\rangle=f_\pi$ and the curvature masses, respectively, are given as
\begin{equation}
f_\pi(m^2+\lambda f_\pi^2)=h,\qquad m_\pi^2=m^2+\lambda f_\pi^2=\frac{h}{f_\pi},\qquad m_\sigma^2=m^2+3\lambda f_\pi^2=m_\pi^2+2\lambda f_\pi^2. \label{eq:su2gap}
\end{equation}
Thus, $\lambda=(m_\sigma^2-m_\pi^2)/(2f_\pi^2)$ and the standard PCAC relation $f_\pi m_\pi^2=h$ \cite{gellmann60,lenaghan00}. In the present normalization, the constituent quark mass is $m_q=g\sigma/2$ where $g\to2g$ for $m_q=g\sigma$. In this truncation, we omit the isovector $a_0$ and isoscalar $\eta$ fields of $U(2)$, which would otherwise complete the $(\sigma, \bm{\pi})$ fields into a full $SO(4)$ vector \cite{gellmann60}.

\subsection{The $SU(3)$ model and larger flavour groups}\label{sec:su3}

For $N_f=3$ the condensates $\sigma_0,\sigma_8$ are rotated to the nonstrange--strange basis
\begin{equation}
\begin{pmatrix}\sigma_x\\ \sigma_y\end{pmatrix}=\frac1{\sqrt3}\begin{pmatrix}\sqrt2&1\\1&-\sqrt2\end{pmatrix}\begin{pmatrix}\sigma_0\\ \sigma_8\end{pmatrix}, \label{eq:xy}
\end{equation}
of an orthogonal, symmetric (hence self-inverse) matrix. Using the explicit representations $T_0 = \mathbf{1}/\sqrt{6}$ and $T_8=\mathrm{diag}(1, 1, -2)/(2\sqrt{3})$, the $(1,1)$ matrix entry evaluates to $\sigma_0/\sqrt{6}+\sigma_8/(2\sqrt{3})=\sigma_x/2$, while the $(3,3)$ entry yields $\sigma_y/\sqrt{2}$ so
\begin{equation}
\langle\Phi\rangle=\mathrm{diag}\bigl(\tfrac{\sigma_x}{2},\tfrac{\sigma_x}{2},\tfrac{\sigma_y}{\sqrt2}\bigr),\qquad \langle H\rangle=\mathrm{diag}\bigl(\tfrac{h_x}{2},\tfrac{h_x}{2},\tfrac{h_y}{\sqrt2}\bigr). 
\label{eq:vev}
\end{equation}
On this background, $\Tr\Phi^\dagger\Phi=\tfrac12(\sigma_x^2+\sigma_y^2)$ and $\Tr(\Phi^\dagger\Phi)^2=\tfrac18\sigma_x^4+\tfrac14\sigma_y^4$, $\det\Phi=\sigma_x^2\sigma_y/(4\sqrt2)$, so
\begin{align}
U_{\rm LSM}(\sigma_x,\sigma_y)=&\tfrac{m^2}{2}(\sigma_x^2+\sigma_y^2)-h_x\sigma_x-h_y\sigma_y-\tfrac{c}{2\sqrt2}\sigma_x^2\sigma_y + \tfrac{2\lambda_1+\lambda_2}{8}\sigma_x^4+\tfrac{\lambda_1}{2}\sigma_x^2\sigma_y^2+\tfrac{\lambda_1+\lambda_2}{4}\sigma_y^4. \label{eq:Uxy}
\end{align}
This closed form was verified against the matrix-form potential Eq.~\eqref{eq:Llsm} evaluated with machine-precision differentiation. The stationarity conditions are expressed as
\begin{align}
m^2 \sigma_x+\tfrac{2\lambda_1+\lambda_2}{2}\sigma_x^3+\lambda_1\sigma_x\sigma_y^2-\tfrac{c}{\sqrt2}\sigma_x\sigma_y-h_x &=0, \label{eq:gapx0}\\
m^2 \sigma_y+\lambda_1\sigma_x^2\sigma_y+(\lambda_1+\lambda_2)\sigma_y^3-\tfrac{c}{2\sqrt2}\sigma_x^2-h_y &=0. \label{eq:gapy0}
\end{align}
From PCAC relation, we obtain $\sigma_x=f_\pi$, $\sigma_y=(2f_K-f_\pi)/\sqrt2$, $h_x=f_\pi m_\pi^2$, and $h_y=\sqrt2f_Km_K^2-f_\pi m_\pi^2/\sqrt2$. The quark--meson coupling $-g\bar\psi T_a(\sigma_a+i\gamma_5\pi_a)\psi$ gives $m_l=g\sigma_x/2$ and $m_s=g\sigma_y/\sqrt2$.

To fix the model parameters, the meson mass matrix is to be derived by differentiating the matrix-form potential. Its entries are exact quadratic polynomials in $(\sigma_x, \sigma_y)$, fitted from fourteen sample points with a relative residual below $10^{-6}$. Imposing the gap equations\footnote{Using the dispersion relation $\partial E/\partial m=m/E$ alongside the thermal occupation functions for a quark of flavor $f$, the mass derivative of the quark grand potential reads
\begin{equation*}\frac{\partial\Omega_{q\bar q}^{(f)}}{\partial m_f} = 6\int\frac{\dd^3k}{(2\pi)^3}\frac{m_f}{E_f}\left(n_q^+ + n_q^-\right),
\end{equation*}
where $n_q^+=(\phi z+2\bar\phi z^2 + z^3)/D_+$ and $D_+=1 + 3\phi z + 3\bar\phi z^2 + z^3$ (with $n_q^-$ and $D_-$ defined analogously under $z\to\bar z$ and $\phi \leftrightarrow \bar\phi$). Differentiation with respect to the Polyakov-loop variables yields $\partial_\phi\ln D_+=3z/D_+$, $\partial_{\bar\phi}\ln D_+=3z^2/D_+$, $\partial_\phi\ln D_-=3\bar z^2/D_-$, and $\partial_{\bar\phi}\ln D_-=3\bar z/D_-$. Collecting all sectors, the stationary gap equation for the non-strange condensate takes the following form"
\begin{equation*}\partial_{\sigma_x}\Omega = \partial_{\sigma_x}U_{\rm LSM} + g,\partial_{m_l}\Omega_{q\bar q}^{(l)} + \sum_{s=\pm}2,\partial_{\sigma_x}M_{l,s}^2,\partial_{M^2}\Omega_{\tilde q} + \kappa\sigma_x\partial_{m_O^2}\Omega_O + \partial_{\sigma_x}\Omega_{\rm mes} = 0,
\end{equation*}
with an analogous condition for the strange condensate $\sigma_y$ and portal coupling $\kappa$. Simultaneously, stationarity with respect to $\phi$ requires that 
\begin{equation*}\partial_\phi\Omega = \partial_\phi U_{\rm Poly} - 6T\sum_f\int\left[\frac{z_f}{D_+} + \frac{\bar z_f^2}{D_-}\right] + 3T\sum\int\left[-\frac{\tilde z}{B_+} + \frac{\bar{\tilde z}^2}{B_-}\right] + \partial_\phi\Omega_O + \partial_\phi\Omega_{\rm mes} = 0,
\end{equation*}
paired with its conjugate equation $\partial_{\bar\phi}\Omega = 0$. To evaluate the color traces systematically, we express the auxiliary polynomials via standard Newton identities. Expanding the fundamental characteristic polynomial $\prod_j(1+zL_{jj}) = 1 + z e_1 + z^2 e_2 + z^3 e_3$ and enforcing $\det L=1$ give $e_2= \Tr L^{-1}=3\bar\phi$, which is in agreement with Eq.~\eqref{eq:detfund}. Applying Newton's identities, namely, $p_1 = e_1$, $p_2 = e_1p_1 - 2e_2$, $p_3 = e_1p_2 - e_2p_1 + 3e_3$, and the recursion $p_k = e_1p_{k-1} - e_2p_{k-2} + e_3p_{k-3}$ for $k \ge 4$, yields the higher winding traces $p_2 = 9\phi^2 - 6\bar\phi$ and $p_3 = 27\phi^3 - 27\phi\bar\phi + 3$. In the adjoint representation, the elementary symmetric polynomials follow recursively from $k e_k = \sum_{i=1}^k (-1)^{i-1} e_{k-i} P_i$ with $e_8 = 1$, ensuring that the $SU(3)$ Haar measure vanishes identically at the $Z_3$ center elements (such as $\phi=\bar\phi=1$).} 
together with the empirical meson masses ($m_\pi, m_K, m_\eta, m_{\eta'}$) yields a system of rank three for $(m^2, \lambda_1, \lambda_2, c)$. One flat direction remains undetermined and is parametrized by $m_\sigma$, defined as the lowest eigenvalue of the isoscalar scalar matrix. Consequently, the couplings $\lambda_2=46.507$ and $c=4785.5\text{ MeV}$ are entirely independent of $m_\sigma$, whereas $(m^2, \lambda_1)$ vary accordingly. The explicit-breaking sources are fixed to $h_x^{1/3}=120.73\text{ MeV}$ and $h_y^{1/3}=336.22\text{ MeV}$, while varying $m_\sigma$ between $400\text{ MeV}$ and $800\text{ MeV}$ gives $\lambda_1=-5.93$ and $13.47$, respectively. These values are fully consistent with Ref.~\cite{schaefer05}.

For $N_f=4$, i.e., $SU(4)$ with a charm condensate $\sigma_c$, the model has 16 scalars and 16 pseudoscalars and $\det\Phi$ is quartic in the fields \cite{tawfik16_su4,diab18_su4,ahmadov24_su4,tawfik25_su4,friesen26_su4}. In a supersymmetric completion the F-term of $W=c\det\hat\Phi$ is then sextic, Section~\ref{sec:susy}. Thus the renormalizable $SU(4)$ quartic structure cannot arise from supersymmetric terms. In this analysis, a full $SU(4)$ loop-corrected fit was not performed. We only present a minimal charm-quark check in Section~\ref{sec:results}.

\section{The Polyakov linear sigma model}\label{sec:plsm}


At $T>0$ Euclidean time is compact, $\tau\in[0,\beta]$ \cite{lebellac00,kapusta06,laine16}. In the Polyakov gauge the temporal gauge field is static and diagonal, and
\begin{equation}
L=\mathcal P\exp\Bigl[i\!\int_0^\beta\!\dd\tau A_4\Bigr]=\mathrm{diag}(e^{i\theta_1},e^{i\theta_2},e^{i\theta_3}),\qquad \sum_j\theta_j=0,\qquad \phi=\tfrac13\Tr L,\ \ \bar\phi=\tfrac13\Tr L^\dagger. \label{eq:poly}
\end{equation}
In pure Yang--Mills theory, the Polyakov loop $\phi$ serves as the order parameter for $Z_3$ center symmetry \cite{polyakov78,svetitsky82}, which is explicitly broken upon introducing fundamental fermions \cite{pisarski00}. At vanishing chemical potential, $\mu = 0$, the mean fields satisfy $\bar{\phi} = \phi^*$, whereas at non-zero chemical potentials, $\mu \neq 0$, $\phi$ and $\bar{\phi}$ become real-valued and independent. For $SU(3)$, $e_1(L)=3\phi$, $e_2(L)=3\bar\phi$, $e_3=1$, so
\begin{equation}
\det(1\pm zL)=1\pm3\phi z+3\bar\phi z^2\pm z^3. \label{eq:detfund}
\end{equation}
The power sums $p_k=\Tr L^k$ obey $p_1=3\phi$, $p_2=9\phi^2-6\bar\phi$, and $p_3=27\phi^3-27\phi\bar\phi+3$. For $k\ge4$, Newton's identity $p_k=3\phi p_{k-1}-3\bar\phi p_{k-2}+p_{k-3}$. For $\bar p_k$, we just $\phi\leftrightarrow\bar\phi$. The adjoint character is given as $\chi_{\rm adj}(L^k)=p_k\bar p_k-1$. The Haar measure is proportional to
\begin{equation}
\mathcal H(\phi,\bar\phi)=1-6\phi\bar\phi+4(\phi^3+\bar\phi^3)-3(\phi\bar\phi)^2, \label{eq:haar}
\end{equation}
which means that it vanishes at the center elements.

The PLSM Lagrangian is constructed as 
\begin{eqnarray}
\mathcal L=\bar\psi[i\gamma^\mu D_\mu-gT_a(\sigma_a+i\gamma_5\pi_a)]\psi+\mathcal L_{\rm LSM}-U_{\rm Poly}, 
\end{eqnarray}
with gluons entering only via the Polyakov loop. Four classes of glue potential have been proposed including the one of this work. They are compared in Tab.~\ref{tab:poly},
\begin{align}
\frac{U_{\rm poly}}{T^4}&=-\frac{b_2(T)}{2}\phi\bar\phi-\frac{b_3}{6}\left(\phi^3+\bar\phi^3\right)+\frac{b_4}{4}\left(\phi\bar\phi\right)^2,\quad b_2(T)=a_0+a_1\tfrac{T_0}{T}+a_2\tfrac{T_0^2}{T^2}+a_3\tfrac{T_0^3}{T^3}, \label{eq:Upoly}\\
\frac{U_{\log}}{T^4}&=-\frac{a(T)}{2}\phi\bar\phi+b(T)\ln\mathcal H,\hspace*{29mm}  a(T)=a_0+a_1\tfrac{T_0}{T}+a_2\tfrac{T_0^2}{T^2},\ b(T)=b_3\tfrac{T_0^3}{T^3}, \label{eq:Ulog}\\
U_{\rm F}&=-bT\Bigl[54e^{-a/T}\phi\bar\phi+\ln\mathcal H\Bigr]. \label{eq:Ufuk}
\end{align}
All four potentials are invariant under the center $Z_3$ symmetry. In practice, enforcing $Z_3$ invariance limits the permissible couplings and terms in each potential to combinations that transform trivially under the symmetry. Depending on the physical setting, such as flavor models, scalar sectors, or effective field theories. This symmetry organizes the parameter space and excludes unwanted terms.

\begin{table}[htb]
\centering
\begin{tabular}{|c|c|c|c|}
\hline \hline
Potential & Haar measure & Parameters \\
\hline \hline
Polynomial Eq.~\eqref{eq:Upoly} & no & $a_0=6.75$, $a_1=-1.95$, $a_2=2.625$, $a_3=-7.44$, $b_3=0.75$, $b_4=7.5$ \\
Logarithmic Eq.~\eqref{eq:Ulog} & yes & $a_0=3.51$, $a_1=-2.47$, $a_2=15.2$, $b_3=-1.75$ \\
Fukushima Eq.~\eqref{eq:Ufuk} & yes & $a$, $b$ fitted to the quenched EoS \\
Calibrated (this work) & yes & $a_0=3.51$ fixed; $T_0,a_1,a_2,b_3$ fitted to lattice thermodynamics, Section~\ref{sec:calib}, \\
\hline
\end{tabular}
\caption{Four Polyakov-loop effective potentials \cite{ratti06,rossner07,fukushima04}, whose parameters are determined, for example, at critical temperature of $T_0 = 270\text{ MeV}$ in the pure-gauge limit. All four potentials have $Z_3$ symmetry.} 
\label{tab:poly}
\end{table}

Stefan--Boltzmann limit can be checked for $T\to\infty$, where the pure-gauge pressure must approach $8\pi^2/45=1.755$. This limit can be estimated in the polynomial form $-U/T^4\to a_0/2+b_3/3-b_4/4=3.375+0.25-1.875$, while in the logarithmic form the parameter $b\to0$ and $-U/T^4\to a_0/2=1.755$.

For thermodynamic consistency let $\varphi$ represent all mean fields. Then with $P=-\Omega(\varphi_*)$ at the global stationary point and $\partial_i\Omega|_{\varphi_*}=0$, the envelope theorem gives
\begin{equation}
s=-\frac{\dd\Omega}{\dd T}=-\frac{\partial\Omega}{\partial T}\Big|_\varphi,\qquad n=-\frac{\partial\Omega}{\partial\mu}\Big|_\varphi, \label{eq:envelope}
\end{equation}
so the explicit $T$-dependence of $a_i(T),b_i(T)$ is automatically included. Providing that all $T$ and $\mu$ dependences including any $T_0(\mu)$ are kept inside $\Omega$ than $\epsilon+P=Ts+n\mu$ holds. This is verified numerically in Section~\ref{sec:num}.

\section{Supersymmetric degrees of freedom}
\label{sec:susy}

\subsection{Field content and symmetry constraints}

In the $\mathcal{N}=1$ formulation of the quark-meson system, each flavor is represented by quark/squark chiral superfields $\hat{Q}=(\tilde{q}_L, q_L, F)$ and $\hat{Q}^c=(\tilde{q}_R^*, q_R^c, F^c)$, as well as a meson superfield $\hat{\Phi}=(\Phi, \psi_\Phi, F_\Phi)$. The vector multiplet $\hat{V}=(A_\mu, \lambda, D)$ provides the gauge fields and the Majorana gluino $\lambda$. Following standard conventions of Ref.~\cite{binetruy09}, the sgluon is excluded from $\hat{V}$ and is instead treated as the color-octet scalar of an adjoint chiral superfield $\hat{\Sigma}=(O^a, \psi_\Sigma^a, F_\Sigma^a)$. Analogous to $\mathcal{N}=2$, SQCD and Dirac-gluino scenarios \cite{fayet78, fox02, binetruy09, kribs12}, $\lambda$ and $\psi_\Sigma$ pair up to construct a Dirac gluino. Furthermore,
\begin{equation}
O^a=\tfrac1{\sqrt2}(S^a+iP^a),\qquad a=1,\dots,8, \label{eq:sgluon}
\end{equation}
is a complex adjoint scalar with scalar and pseudoscalar parts \cite{benakli10}. The identification that makes the embedding natural is that $\Phi=T_a(\sigma_a+i\pi_a)$ is already a complex matrix. The scalar and pseudoscalar nonets are the real and imaginary parts of the scalar component of one chiral multiplet. The new states are the fermionic partners (mesinos). The following symmetries constrain the effective couplings:
\begin{itemize}
\item \emph{Colour gauge invariance} $SU(3)_c$: $\tq$ in the fundamental and $O^a$ is the adjoint.
\item \emph{Chiral flavour invariance} $U(3)_L\times U(3)_R$ for portals to $\Phi$: $\tq_L\to U_L\tq_L$ and $\tq_R\to U_R\tq_R$ with $O^a$ a flavour singlet.
\item \emph{Baryon number}: $\tq$ carries $B=1/3$, while $O^a$ none.
\item \emph{Center symmetry $Z_3$:} The adjoint $O^a$ preserves it, while the fundamental $\tq$ breaks it explicitly, as quarks do. Consequently sgluons cannot induce $\langle\phi\rangle\neq0$ and act as additional glue, while squarks act as bosonic partners of the quark.
\item \emph{R-parity}: squarks are R-odd and enter pairwise, while sgluons are R-even.
\end{itemize}

\subsection{Constraints of holomorphy on the meson potential}
\label{sec:holo}

The superpotential $W$ must be holomorphic in $\hat\Phi$ that the superpotential $W(\hat\Phi)$ depends analytically and exclusively on the chiral superfield $\hat\Phi$ itself but does not depend on its Hermitian conjugate $\hat\Phi^\dagger$ or any spacetime derivatives. The only $SU(N_f)_L\times SU(N_f)_R$-invariant holomorphic function is $\det\hat\Phi$, so $\mathrm{Tr}\,\Phi^\dagger\Phi$ cannot appear in $W$. It lives in the K\"ahler potential $K=\Tr\Phi^\dagger\Phi$, which gives the kinetic term Eq.~\eqref{eq:kin} \cite{binetruy09,wess92}. With $W_\Phi=c_W\det\hat\Phi$, the F-term potential is $|\partial W/\partial\Phi|^2=|c_W|^2\,|\mathrm{cof}\,\Phi|^2$. Thus, in terms of the eigenvalues $s_i$ of $\Phi^\dagger\Phi$, we have
\begin{equation}
V_F^{(N)}=|c_W|^2\,e_{N-1}(s_1,\dots,s_N), \label{eq:VFN}
\end{equation}
with $e_k$ are the elementary symmetric polynomials. For $N=3$, $e_2=\tfrac12[(\Tr\Phi^\dagger\Phi)^2-\Tr(\Phi^\dagger\Phi)^2]$, which is exactly the LSM quartic with
\begin{equation}
\lambda_1=-\lambda_2=\tfrac12|c_W|^2. \label{eq:lamrel}
\end{equation} 
For check on the $(\sigma_x,\sigma_y)$ background, we have $V_F=|c_W|^2[\sigma_x^4/16+\sigma_x^2\sigma_y^2/4]$, while Eq.~\eqref{eq:Uxy} with $\lambda_2=-\lambda_1=-|c_W|^2/2$ gives $(2\lambda_1+\lambda_2)/8=|c_W|^2/16$, $\lambda_1/2=|c_W|^2/4$, and $(\lambda_1+\lambda_2)/4=0$, for  the $\sigma_x^4$, $\sigma_x^2\sigma_y^2$, and $\sigma_y^4$ coefficients. Thus the supersymmetric part of the potential does not allow independent $\lambda_{1,2}$ for $N_f=3$. The $m^2$ term should be negative to break chiral symmetry. The 't~Hooft term is the soft trilinear $A$-term $c_A(\det\phi+\text{h.c.})$. The explicit source must be soft  supersymmetry breaking. The fitted values $\lambda_1=-4.4$, $\lambda_2=46.5$ of the calibrated approach seems to violate Eq.~\eqref{eq:lamrel}, i.e., our numerical model is not exactly supersymmetric in its meson sector. Let us now discuss various configurations:
\begin{itemize}
\item{$N_f=2$:} Here $\det\hat\Phi$ is quadratic and gives only a mass term. A quartic arises with a gauge-singlet Lagrange-multiplier superfield $\hat S$. Writing the complex $SO(4)$ vector $Z=(x+iy)/\sqrt2$ with $x=(\sigma,\bm\pi)$ and $y=(\eta,\bm a_0)$ leads to
\begin{equation}
W=\kappa\,\hat S\,(Z\!\cdot\!Z-\mu^2)\ \Rightarrow\ V_F=\kappa^2|Z\!\cdot\!Z-\mu^2|^2+4\kappa^2|S|^2|Z|^2, \label{eq:Wsu2}
\end{equation}
where $\kappa$ is portal coupling.
On $y=0$, Eq.~\eqref{eq:Wsu2} is $\lambda(\sigma^2+\bm\pi^2-v^2)^2/4$ with $\lambda=\kappa^2$, $v^2=2\mu^2$, and the ${\mathcal O}(4)$ is Mexican hat. Around $x=(v,\bm0)$ one finds that  $m_\sigma^2=m_\eta^2=2\kappa^2v^2$ equals to the fermion mass$^2$ from $\psi_S$--$\psi_Z$. Mixing $M=2\kappa|Z|=\sqrt2\kappa v$ confirms supermultiplet degeneracy. The pions are massless Goldstones and their partners $\bm a_0$ are massless K\"ahler partners of the complexified vacuum manifold $Z\cdot Z=\mu^2$ \cite{shore89,bando84,bando84b}.
\item{$N_f=4$:} Because $\det\hat{\Phi}$ is quartic, the associated $F$-term potential $V_F^{(4)}=\vert{}c_W\vert{}^2 e_3(s)$ becomes sextic, governed by a coupling $c_W$ of mass dimension $-1$. This highlights that the $SU(4)$ framework is inherently an effective field theory, where conventional quartic couplings $\lambda_{1,2}$ fail to arise from pure supersymmetric interactions.
\item{Exact SQCD results:} For $N_c=3$ SQCD, the infrared dynamics take three distinct forms depending on the number of flavors $N_f$ \cite{ads84,seiberg94,intriligator96}:
\begin{itemize}
\item $N_f=2$: An Affleck–Dine–Seiberg (ADS) runaway superpotential, $W \propto \Lambda^7 / \det M$.
\item $N_f=3$: A deformed moduli space with the quantum constraint $\det M-B\tilde{B}=\Lambda^6$, equivalently written via a Lagrange multiplier $S$ as $W=S(\det\Phi-B\tilde{B}-\Lambda^6)$.
\item $N_f=4$: Smooth $s$-confinement described by the superpotential $W=\Lambda^{-5}(BM\tilde{B}- \det M)$.
\end{itemize}
While these exact results pin down the holomorphic dependence on $\det\Phi$ and the baryonic superfields $B$ and $\tilde{B}$, they do not fix the details of our hybrid constituent-quark setup. Consequently, we use them only as qualitative guidance in the numerical model rather than as strict constraints.
\end{itemize}

\subsection{Squarks}

For each flavour we introduce complex colour-triplet scalars $\tq_{f,L},\tq_{f,R}$ with
\begin{equation}
\mathcal L_{\tilde q}=\sum_f\bigl(|D_\mu\tq_{f,L}|^2+|D_\mu\tq_{f,R}|^2\bigr)-V_{\tilde q},\qquad D_\mu=\partial_\mu-ig_sA^a_\mu T^a. \label{eq:Lsq}
\end{equation}
From $W\supset g\hat Q^c\hat\Phi\hat Q$, eliminating the auxiliary fields gives the F-term potential $V^F_{\tilde q}=g^2[\tq_R^\dagger\Phi^\dagger\Phi\,\tq_R+\tq_L^\dagger\Phi\Phi^\dagger\,\tq_L]$. Each term is invariant since $\Phi^\dagger\Phi\to U_R\Phi^\dagger\Phi U_R^\dagger$ acts on $\tq_R$. Also eliminating the auxiliary fields gives the soft sector $V^{\rm soft}_{\tilde q}=m_0^2(|\tq_L|^2+|\tq_R|^2)+[A_qg\,\tq_L^\dagger\Phi\tq_R+\text{h.c.}]$. On the background Eq.~\eqref{eq:vev} with $m_f=g\langle\Phi\rangle_{ff}$, the squared-mass matrix in the $(\tq_{f,L},\tq_{f,R})$ basis is written as:
\begin{eqnarray}
M_{f,\pm}^2 &=& \begin{pmatrix}m_0^2+m_f^2&A_qm_f\\A_qm_f&m_0^2+m_f^2\end{pmatrix} =m_0^2+m_f^2\pm A_qm_f. \label{eq:Mf}
\end{eqnarray}
The squark spectrum therefore responds to chiral restoration through $m_f(\sigma_{x,y})$. In the numerical analyses as shall be elaborated only $A_q=0$ is used and the squarks form a free gas. The D-term quartic $g_s^2(\tq^\dagger T^a\tq)^2/2$, the F-term cross terms with $\det\Phi$, thermal gauge-induced masses $\Pi_{\tilde q}\propto g_s^2T^2$, and condensates are neglected. They enter at two loops through Hartree self-energies \cite{cornwall74,berges04}. Single-squark condensation $\langle\tilde q\rangle\neq0$ would break $SU(3)_c$ \cite{casas96}, so we set $\langle\tilde q\rangle=0$. We therefore restrict the analyses to $M_{f,-}>|\mu_q|$ and flag any violation as an instability of the free-gas approximation.

\subsection{Sgluons}

The dynamics of the sgluon sector are governed by the Lagrangian
\begin{equation}
\mathcal{L}_O = (D\mu O)^{a\dagger}(D^\mu O)^a - V_O,
\label{eq:LO}
\end{equation}
where the covariant derivative and scalar potential are defined as
\begin{equation}
D_\mu O^a = \partial_\mu O^a + g_s f^{abc} A^b_\mu O^c, \qquad
V_O = m_{O,0}^2 O^{a\dagger} O^a + \kappa,\operatorname{Tr}(\Phi^\dagger\Phi) O^{a\dagger} O^a + \frac{\lambda_O}{4}(O^{a\dagger} O^a)^2.
\end{equation}
Evaluating the potential on the chiral mean-field background shifts the bare sgluon mass parameter $m_{O,0}^2$, yielding an effective squared mass of
\begin{equation}
m_O^2 = m_{O,0}^2 + \frac{\kappa}{2}(\sigma_x^2 + \sigma_y^2).
\end{equation}

While the $\mathcal{N}=2$ superpotential term $W \supset \sqrt{2} g_s \hat{Q}^c \hat{\Sigma} \hat{Q}$ inherently generates a squark-sgluon quartic interaction $2g_s^2 \tilde{q}^\dagger T^a T^b \tilde{q} \, O^{a\dagger} O^b$ besides standard Yukawa couplings, our numerical analysis seems to omit these contributions as well as the sgluon self-coupling $\lambda_O$. Instead, we retain only the chirally invariant portal coupling $\kappa$, which couples the sgluon fields $O^a$ directly to the meson matrix as an effective term rather than a strict consequence of supersymmetry. Finally, in the presence of the Polyakov-loop background, the Matsubara frequencies of $O^a$ are shifted by $w \cdot \theta \, T$ for each adjoint weight $w$.

\section{Thermodynamics}\label{sec:thermo}

The extended effective potential reads
\begin{equation}
\Omega=U_{\rm LSM}+U_{\rm Poly}+\Omega_{q\bar q}+\Omega_{\rm mes}+\Omega_{\tilde q}+\Omega_O\equiv\Omega(T,\mu;\sigma_x,\sigma_y,\phi,\bar\phi). \label{eq:Ueff}
\end{equation}
Note that the $\overline{\rm MS}$ vacuum term per field is given as  \cite{coleman73}
\begin{equation}
\Omega^{\rm vac}=\frac{1}{64\pi^2}\sum_i(-1)^{F_i}n_iM_i^4 \left[\ln(M_i^2/Q^2)-\tfrac32\right],
\end{equation}
In this regard, $\Omega^{\rm vac}$ cancels for unbroken supersymmetry but with soft breaking it does not. It is omitted in the numerical approach in Section~\ref{sec:reg}.

The grand partition function is
\begin{equation}
\mathcal Z=\Tr e^{-\beta(\hat H-\mu\hat N)}=\int\mathcal D\varphi\,\mathcal D\bar\psi\mathcal D\psi\,\mathcal D\tq^\dagger\mathcal D\tq\,\mathcal DO^\dagger\mathcal DO\,\exp\left[-\!\int_0^\beta\!\dd\tau\!\int\!\dd^3x\,\mathcal L_E\right], \label{eq:Z}
\end{equation}
with periodic bosons and antiperiodic fermions. Thermodynamically, the potential is derived as $\Omega=-T\ln\mathcal Z/V$. In the mean-field approximation, the mesons are constants and the remaining fields appear quadratically, so a complex boson gives $\det^{-1}$ and a Dirac fermion $\det$ of an operator with eigenvalues $(\omega_n+\theta_jT-i\mu)^2+E^2$ for each colour phase $e^{i\theta_j}$ of $L$, with $\omega_n=2n\pi T$ (bosons) or $(2n+1)\pi T$ (fermions). For Matsubara sums let us propose a \emph{Lemma} that for real $x>0$ and complex $\varphi$, we have 
\begin{align}
\prod_{n\in\mathbb Z}\bigl[(2\pi n+\varphi)^2+x^2\bigr]&\widehat{=}\,2(\cosh x-\cos\varphi)=e^x(1-e^{-x+i\varphi})(1-e^{-x-i\varphi}), \label{eq:lemB}\\
\prod_{n\in\mathbb Z}\bigl[((2n+1)\pi+\varphi)^2+x^2\bigr]&\widehat{=}\,2(\cosh x+\cos\varphi)=e^x(1+e^{-x+i\varphi})(1+e^{-x-i\varphi}), \label{eq:lemF}
\end{align}
up to a $\varphi$- and $x$-independent constant after zeta regularization. To prove this Lemma, we write the factor as $(2\pi n+\varphi+ix)(2\pi n+\varphi-ix)$ and use $\sin(z/2)=\tfrac z2\prod_{n\ge1}(1-z^2/4\pi^2n^2)$ with $z=\varphi\pm ix$, and $\sin\frac{\varphi+ix}{2}\sin\frac{\varphi-ix}{2}=\tfrac12(\cosh x-\cos\varphi)$. By shifting $\varphi\to\varphi+\pi$, the fermionic case is obtained, then $2\cosh x\mp2\cos\varphi=e^x\mp e^{i\varphi}\mp e^{-i\varphi}+e^{-x}$ gives the factorized forms. With $\varphi=\theta_j-i\mu/T$, the fermionic identity gives $\sum_n\ln[(\omega_n+\theta_jT-i\mu)^2+E^2]=E/T+\ln(1+e^{-(E-\mu)/T+i\theta_j})+\ln(1+e^{-(E+\mu)/T-i\theta_j})+\text{const}$. The bosonic one is the same but with flipping the sign in $(1+\cdots)$ to $(1-\cdots)$. The upper branch (particles) couples to $L$, while the lower branch (antiparticles) couples to $L^\dagger$.

\subsection{Quark, squark, sgluon and meson sectors}

Summing over colour with Eq.~ \eqref{eq:detfund} and over spin, a Dirac spinor contributes $[\cdots]^2$ leads to
\begin{equation}
\Omega_{q\bar q}=-2N_c\sum_f\!\int\!\frac{\dd^3k}{(2\pi)^3}E_f-2T\sum_f\!\int\!\frac{\dd^3k}{(2\pi)^3}\Bigl\{\ln\det(1+z_fL)+\ln\det(1+\bar z_fL^\dagger)\Bigr\}, \label{eq:Omq}
\end{equation}
with $z_f=e^{-(E_f-\mu)/T}$, $\bar z_f=e^{-(E_f+\mu)/T}$. Note that the first term is the vacuum part. Each complex colour-triplet boson of mass $M$ gives $\ln\mathcal Z=-\sum_{n,j}\ln[\cdots]$, so with $E_{f,s}=\sqrt{k^2+M_{f,s}^2}$, we get 
\begin{equation}
\Omega_{\tilde q}=T\sum_{f,s=\pm}\!\int\!\frac{\dd^3k}{(2\pi)^3}\Bigl\{\ln[1-3\phi\tilde z+3\bar\phi\tilde z^2-\tilde z^3]+\ln[1-3\bar\phi\bar{\tilde z}+3\phi\bar{\tilde z}^2-\bar{\tilde z}^3]\Bigr\}, 
\label{eq:Omsq}
\end{equation}
where $\tilde z=e^{-(E-\mu)/T}$ and $\mu=\mu_q$. As discussed squarks carry $B=1/3$. Thus there are $2N_f$ such species. The adjoint is real, so particles and antiparticles both couple to $L_{\rm adj}$ and there is no chemical potential. Using $\ln\det_{\rm adj}(1-zL)=-\sum_kz^k\chi_{\rm adj}(L^k)/k$ and $\int\frac{\dd^3q}{(2\pi)^3}e^{-kE_q/T}=\frac{m^2T}{2\pi^2k}K_2(km/T)$, then
\begin{equation}
\Omega_O=-\frac{m_O^2T^2}{\pi^2}\sum_{k=1}^\infty\frac{K_2(km_O/T)}{k^2}\bigl[p_k\bar p_k-1\bigr]. \label{eq:OmO}
\end{equation}
For check, let $L={\mathbf 1}$ and $p_k\bar p_k-1=8$. Then for $m\to0$, $K_2(x)\to2/x^2$ gives $-\tfrac{16\pi^2}{90}T^4$, i.e., the pressure of 16 real massless bosons. In the numerical implementation, the adjoint determinant is evaluated directly as a polynomial in $z$ with coefficients from Newton identities \cite{macdonald95}, $\det_{\rm adj}(1-zL)=\sum_{j=0}^{8}(-z)^j e_j^{\rm adj}$, where $e_j^{\rm adj}$ are the elementary symmetric polynomials in the $8$ adjoint eigenvalues of $L$.

Additionally, the applied numerical approach accounts for finite-temperature quantum fluctuations via thermal loops of the pseudo-Goldstone octet,
\begin{equation}
\Omega_{\rm mes}=w(\phi,\bar\phi)\sum_{i\in\{\pi,K,\eta\}}n_i\,T\!\int\!\frac{\dd^3k}{(2\pi)^3}\ln\bigl(1-e^{-E_i/T}\bigr),\qquad w=1-\phi\bar\phi, 
\label{eq:Ommes}
\end{equation}
with tree-level curvature masses $M_i(\sigma_x,\sigma_y)$. The weight $w$ is a statistical-confinement-like suppression of colour-singlet hadronic states above $T_c$ \cite{megias06}. We emphasize that this is just a modelling assumption not a mathematical derivation.

\subsection{Observables}

With $P=-\Omega(\varphi_*)$, where vacuum term is subtracted, and with Eq.~\eqref{eq:envelope} and thermodynamics identities\footnote{With $\Omega(T,\mu)$ at the stationary point, $P=-\Omega$, $s=-\partial_T\Omega$, $n=-\partial_\mu\Omega$, $\epsilon=Ts+\mu n-P$, $\dd\epsilon=T\dd s+\mu\dd n$, $C_V=\partial_T\epsilon|_{n}$. At $\mu=0$, $\dd\epsilon=T\dd s$ so $c_s^2=\dd P/\dd\epsilon=s/(T\,\dd s/\dd T)$. Consistency requires the Maxwell relation $\partial s/\partial\mu=\partial n/\partial T$, verified numerically.}, we find that
\begin{align}
&s=-\frac{\partial\Omega}{\partial T},\qquad\, n_q=-\frac{\partial\Omega}{\partial\mu_q},\qquad\;\; \epsilon=-P+Ts+n_q\mu_q,\nonumber\\
&\Delta=\frac{\epsilon-3P}{T^4},\quad C_V=T\Bigl(\frac{\partial s}{\partial T}\Bigr)_\mu,\quad c_s^2=\Bigl(\frac{\partial P}{\partial\epsilon}\Bigr)_{s/n}. \label{eq:obs}
\end{align}
Note that at $\mu=0$, $c_s^2=s/C_V$. For susceptibilities, for instance, up to fourth order, let $F(\mu)=\Omega(T,\mu;\varphi_*(\mu))$ and $D=\partial_\mu+\varphi_i'\partial_i$, the total derivative along the stationary solution\footnote{Define $D=\partial_\mu+\varphi_i'\partial_i$. Then for any $f(\mu,\varphi)$, 
\begin{eqnarray*}
Df &=& f_\mu+\varphi_i'f_i,  \\ D^2f &=& f_{\mu\mu}+2\varphi_i'f_{\mu i}+\varphi_i'\varphi_j'f_{ij}+\varphi_i''f_i,  \\ 
D^3f &=& f_{\mu\mu\mu}+3\varphi_i'f_{\mu\mu i}+3\varphi_i'\varphi_j'f_{\mu ij}+\varphi_i'\varphi_j'\varphi_k'f_{ijk}+3\varphi_i''f_{\mu i}+3\varphi_i'\varphi_j''f_{ij}+\varphi_i'''f_i. 
\end{eqnarray*}
Stationarity $\Omega_i=0$ implies $D^m\Omega_i=0$, which determines $\varphi_i^{(m)}$, and $F^{(n)}=D^{n-1}\Omega_\mu$ gives Eqs.~\eqref{eq:F2}--\eqref{eq:F4}. Note that for massless quarks $\Omega=-\frac{N_cN_f}{12\pi^2}(\ldots)$ gives $\chi_2^q/T^2=N_cN_f/3$ and $\chi_4^q=2N_cN_f/\pi^2$, whence $\chi_4^B/\chi_2^B=2/(3\pi^2)$.}. Then $F'=\Omega_\mu$ and $F^{(n)}=D^{n-1}\Omega_\mu$. The derivatives of $\varphi_*$ follow from $D^m\Omega_i=0$. With the Hessian $H_{ij}=\Omega_{ij}$, where indices after $\mu$ denote partial derivatives, we get
\begin{align}
\varphi_i'&=-(H^{-1})_{ij}\Omega_{j\mu}, \label{eq:phip}\\
F''&=\Omega_{\mu\mu}-\Omega_{\mu i}(H^{-1})_{ij}\Omega_{j\mu}, \label{eq:F2}\\
F'''&=\Omega_{\mu\mu\mu}+2\Omega_{\mu\mu i}\varphi_i'+\Omega_{\mu ij}\varphi_i'\varphi_j'+\Omega_{\mu i}\varphi_i'', \label{eq:F3}\\
F''''&=\Omega_{\mu\mu\mu\mu}+3\Omega_{\mu\mu\mu i}\varphi_i'+3\Omega_{\mu\mu ij}\varphi_i'\varphi_j'+\Omega_{\mu ijk}\varphi_i'\varphi_j'\varphi_k'+3\Omega_{\mu\mu i}\varphi_i''+3\Omega_{\mu ij}\varphi_i'\varphi_j''+\Omega_{\mu i}\varphi_i''', \label{eq:F4}
\end{align}
where $\varphi''$ and $\varphi'''$ solve $D^2\Omega_i=0$ and $D^3\Omega_i=0$, respectively, so
\begin{align}
\varphi_i''&=-(H^{-1})_{ij}\bigl[\Omega_{j\mu\mu}+2\Omega_{j\mu k}\varphi_k'+\Omega_{jkl}\varphi_k'\varphi_l'\bigr],\nonumber\\
\varphi_i'''&=-(H^{-1})_{ij}\bigl[\Omega_{j\mu\mu\mu}+3\Omega_{j\mu\mu k}\varphi_k'+3\Omega_{j\mu kl}\varphi_k'\varphi_l'+\Omega_{jklm}\varphi_k'\varphi_l'\varphi_m'+3\Omega_{j\mu k}\varphi_k''+3\Omega_{jkl}\varphi_k'\varphi_l''\bigr]. \label{eq:phipp}
\end{align}
These follow from $D^2f=f_{\mu\mu}+2\varphi_i'f_{\mu i}+\varphi_i'\varphi_j'f_{ij}+\varphi_i''f_i$ and the analogous $D^3f$. Because $\mu_q=\mu_B/3$ and $\chi_n=\partial^n(P/T^4)/\partial(\mu/T)^n$, the quark-number susceptibilities are
\begin{equation}
\chi_n^q=-T^{n-4}F^{(n)}(\mu_q),\qquad \chi_n^B=\chi_n^q/3^n,\quad n=2,3,4. \label{eq:chin}
\end{equation}
For massless quarks $\chi_2^B/T^2=1/3$ and $\chi_4^B/\chi_2^B=2/(3\pi^2)$. In practice $\chi_2$ is evaluated from Eq.~\eqref{eq:F2} with automatic differentiation, and $\chi_4$ from a central finite difference of $F''(\mu)$ with the stationary solution re-solved at $\pm\delta\mu$. The analytic expressions above serve as the cross-check.

\section{Phase structure and the critical endpoint}
\label{sec:phase}

Let us start with stationary points, saddle structure, and stability. The mean fields $\varphi=(\sigma_x,\sigma_y,\phi,\bar\phi)$ solve
\begin{equation}
g_i(\varphi;T,\mu)\equiv\partial\Omega/\partial\varphi_i=0. \label{eq:stat}
\end{equation}
Since $\phi,\bar\phi$ parametrize a complex-conjugate pair treated as independent real variables, the physical solution is a saddle of $\Omega$ along $\phi-\bar\phi$ and a minimum along the other directions. We verified numerically that every converged solution has exactly one large negative Hessian eigenvalue, and define a stable solution by
\begin{equation}
n_-(H)=1, \label{eq:stable}
\end{equation}
with the physical one being, among several stable solutions at the same $(T,\mu)$, that with the lowest $\Omega$. Alternatively, minimizing over independent $\phi$ and $\bar\phi$ would not be fully correct. It drives $\bar\phi\to1$ into an unphysical region which was observed explicitly. At $\mu=0$, the reduction $\phi=\bar\phi=r$ is exact and the problem is a minimization over $(\sigma_x,\sigma_y,r)$.


At fixed $T$, a chiral-broken branch is tracked upward in $\mu$ from $\mu=0$ and a restored branch downward in $\mu$ from a large value, each with the stability filter Eq.~\eqref{eq:stable}. Coexistence at $(T,\mu)$ means both converge to stable solutions with $|\sigma_x^{\rm b}-\sigma_x^{\rm r}|>1$~MeV. The first-order line $T_1(\mu)$ is where $\Omega_{\rm b}=\Omega_{\rm r}$. This path-independent construction replaced an up/down temperature sweep that produced spurious hysteresis by branch jumping. Near the CEP, the Landau expansion gives the mean-field relations \cite{landau80,halasz98}
\begin{equation}
(\Delta\sigma_x)^2\propto\mu-\mu_E,\qquad \Delta T_{\rm spinodal}\propto(\mu-\mu_E)^{3/2},\qquad \chi_2^q\propto|\mu-\mu_E|^{-2/3}\ \ (T=T_E), \label{eq:mfscaling}
\end{equation}
used both to extrapolate the tip from coexistence data and as a solver check.


At the CEP, the Hessian acquires a soft eigenvalue and the cubic coupling along it vanishes. Let $\lambda_s$ be the eigenvalue of $H$ closest to zero, excluding the large negative saddle eigenvalue, with unit eigenvector $v$, then $(T_E,\mu_E,\varphi_E)$ solves
\begin{align}
g_i(\varphi;T,\mu)&=0\quad(i=1,\dots,4), \label{eq:cep1}\\
\lambda_s(\varphi;T,\mu)&=0, \label{eq:cep2}\\
\frac{\dd^3}{\dd t^3}\Omega(\varphi+t\,v;T,\mu)\Big|_{t=0}&=0. \label{eq:cep3}
\end{align}
These are six equations for six unknowns, which can be solved by Levenberg--Marquardt \cite{levenberg44,marquardt63} with automatic-differentiation derivatives. Selecting the smallest modulus defines where CEP occurs. In this regard, a selection of, for instance, the smallest positive eigenvalue seem to converge to a different mode.


Searching for CEP would be meaningless if the baseline transition is already first-order at $\mu=0$. To verify the nature of the zero-chemical-potential transition, we initially tested two standard diagnostics, both of which failed validation against known benchmarks and were consequently disregarded. First, relying solely on the maximal slope $\vert{}\mathrm{d}\sigma_x/\mathrm{d}T\vert{}$ proved insufficient, as it cannot reliably distinguish a sharp crossover from a true discontinuity. Second, a two-starting-point bistability test failed to identify a known first-order benchmark. Specifically, the parameter set $m_\sigma=400~\text{MeV}$ with mesonic fluctuations excluded. The adopted procedure sweeps $T$ with step $\delta$ and records
\begin{equation}
\Delta(\delta)=\max_T|\sigma_x(T+\delta)-\sigma_x(T)|,\qquad R=\Delta(0.02~\mathrm{MeV})/\Delta(0.5~\mathrm{MeV}). \label{eq:Rtest}
\end{equation}
For an analytic crossover, we expect $\Delta\propto\delta$ and $R\to0.04$. That is a discontinuity which does not shrink with $\delta$. To check for validation, we evaluate that for $m_\sigma=800$~MeV without mesons. This should give $\Delta=1.53/0.31/0.06$~MeV for $\delta=0.5/0.1/0.02$~MeV, where $R=0.039$, i.e., crossover.  On the other hand, for $m_\sigma=400$~MeV without mesons the estimation gives $20.5/10.1/2.9$~MeV and $R=0.14$, i.e., first order. The pseudocritical temperature is the position of $\max|\dd\sigma_x/\dd T|$ by cubic spline, and similarly for $\sigma_y$ and $\phi$. The same criterion locates the endpoint of the thermal-only $SU(2)$ model below by bracketing between $R<0.1$ and $R>0.1$.

\section{Numerical results}\label{sec:results}

\begin{table}[htb]
\centering
\footnotesize
\resizebox{\textwidth}{!}{%
\begin{tabular}{| c | c | c | c | c |}
\hline \hline
Variant & window, $N$ & $\chi^2$ & $T_\chi$ [MeV] & first order? \\
\hline \hline
Mean-field PLSM, $m_\sigma=600$, $T_0=187$ & 120--280, 42 & $\sim10^4$ & 171 & \textbf{no} but crossover \\
+ 18 meson loops, $(1-\phi\bar\phi)^3$, $m_\sigma=400$ & 120--280, 42 & 2285 & 164 & \textbf{yes} \\
+ vacuum term $Q=250$--$300$ MeV & 120--280, 42 & 7500--8800 & 188--218 & \textbf{no} but crossover \\
+ vacuum term + 18 meson loops & 120--280, 42 & 2050--3000 & -- & \textbf{yes} \\
+ Hartree-resummed thermal meson mass & 120-280, 42 & 4139 & 164 & \textbf{yes} \\
Calibrated $U_{\rm Poly}$, $m_\sigma=800$, no mesons & 170--280, 56 & 34 & 193 & \textbf{no} but crossover \\
Calibrated $U_{\rm Poly}$, $m_\sigma=600$, no mesons & 160--280, 62 & 497 & 163 & \textbf{no} but borderline \\ \hline \hline \hline
\textbf{Applied approach} Octet loops, calibrated, $m_\sigma=450$ & 130--280, 76 & \textbf{257} & \textbf{159} & crossover ($R=0.043$) \\
\hline \hline
\end{tabular}}
\caption{Variants of applied menchmark at $\mu=0$. $\chi^2$ manifests the distance to the measure. $T_\chi$ stands for resulting chiral critical temperatire.  Whether ''first order?'' questions the type of phase transition.}\label{tab:evol}
\end{table}

We designed the numerical analyses in this paper to establish a defensible benchmark. Because of this, the sequence of attempts itself is an important part of the results (see Tab.~\ref{tab:evol}), with $\chi^2$ serving as our distance measure. Three lessons follow from the applied benchmark. First, across every variant we tested, including setups with the vacuum term and Hartree resummation \cite{Dolan:1973qd}, the one-loop thermal contribution of all eighteen mesons (with field-dependent tree-level masses) drives a first-order transition at $\mu=0$. Notably, the thermal-mass tensor stays proportional to the identity within a negligible $10^{-8}$ deviation, yielding $\kappa_{th}\simeq900$--$1650$. Second, large best-fit $T_0$ ($220$--$240$~MeV) strengthens the Polyakov transition and drags the chiral one to first order through the quark coupling. Third, while a calibrated Polyakov potential fits the deconfined window remarkably well ($\chi^2=34$ for $56$ data points at $m_\sigma=800$~MeV), it pushes the crossover temperature up to $T_\chi\simeq193$~MeV. That happens because $T_\chi$ is controlled by the quark--meson sector which is essentially set by $m_\sigma$ rather than by the gluonic potential. We note that lowering $m_\sigma$ lowers $T_\chi$ but steepens the transition. Including only the pseudo-Goldstone octet keeps it analytic down to $m_\sigma=450$~MeV with $T_\chi=158$~MeV.

Details about the numerical analyses at $\mu=0$, concretely for chiral and deconfinement observables, are introduced in Section~\ref{sec:obsrvl}.

\subsection{Numerical analyses at $\mu=0$: Chiral and deconfinement observables}
\label{sec:obsrvl}

\begin{figure}[htb]
\centering
\includegraphics[width=\textwidth]{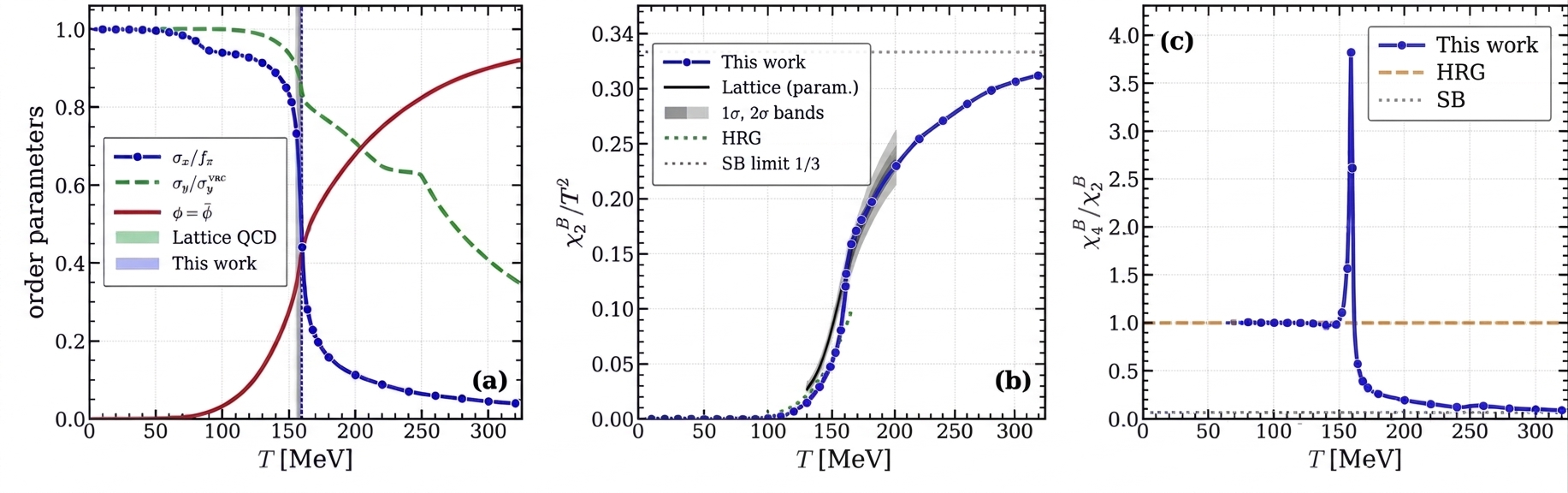}
\caption{Results obtained from numerical analyses at $\mu=0$. Left panel presents the thermal evolution of light condensate $\sigma_x/f_{\pi}$ (blue solid with circles), strange condensate $\sigma_y/\sigma_y^{\text{vc}}$ (dashed curved), and $\phi=\bar\phi$ (red solid). Vertical band marks lattice QCD $T_{pc}\simeq 155-158$~MeV. Middle panel illustrates $\chi_2^B/T^2$ as a function of temperature compared with the parametrization utilized in Ref.~\cite{shah2026locating} with its $1$ and $2\sigma$ band and  16-multiplet HRG results (dashed curved). Stefan--Boltzmann limit $1/3$ is shown as a horizontal dotted line. Right panel depicts the temperature dependence of $\chi_4^B/\chi_2^B$. Dotted line stands for Stefan--Boltzmann $2/3\pi^2$. Dashed line represents HRG-like value typically one).}
\label{fig:orderB}
\end{figure}

In left panel of Fig.~\ref{fig:orderB}, we find that chiral symmetry is nearly intact in its vacuum form up to $\simeq100$~MeV that $\sigma_x/f_{\pi}=0.94$ (blue solid curve with circles) at $100$~MeV, $0.83$ at $150$~MeV, and the Polyakov loop (red solid curved) is negligible that $\phi=0.03$ at $100$~MeV. The light condensate then drops by a factor $\simeq3$ within a few MeV, namely, $0.65$ at $158$, $0.44$ at $160$, and $0.21$ at $170$~MeV, while $\phi$ rises from $0.27$ at $150$~MeV to $0.53$ at $170$~MeV. Both quantities continue smoothly toward $\sigma_x/f_\pi=0.04$ and $\phi=0.92$ at $325$~MeV, without $\phi$ reaching $1$. As expected from the heavier strange quark, the strange condensate (dashed curve) is much more robust. It is still at $0.88$ of its vacuum value at $158$~MeV and falls only to $0.34$ at $325$~MeV. We find that the three transitions are centred at the same temperature, $159.4$~MeV, with FWHM widths of $2.4$ (light), $2.9$ (strange) and $4.0$~MeV (Polyakov loop). This coincidence is partly built in, since the Polyakov-loop calibration was tied to $T_\chi$, Eq.~\eqref{eq:calibobj}, and the quark--meson coupling makes $\sigma$ and $\phi$ respond to each other. Therefore, the coincidence in the three panel seem to be an independent prediction. The transition is, however, much narrower than the lattice crossover. It is slightly above the lattice range (vertical band). A weak secondary structure near $250$~MeV in $\sigma_y$ (right panel) was not analyzed. It may be related to the thermal activation of the strange quark.

In the middle panel the scaled susceptibility $\chi_2^B/T^2$ is analyzed as a function of temperature $T$. It is obvious that below $\simeq~110$~MeV the model susceptibility is negligible, $\chi_2^B/T^2=1.1\times10^{-3}$ at $100$~MeV, because baryon number is carried by three-quark clusters weighted by $\phi^{\,}\!\ll1$. In the region $120$--$152$~MeV the model lies below the lattice parametrization by a factor $0.44$--$0.71$, which is where the missing baryon resonances matter. The 16-multiplet HRG is closer (a factor $0.64$--$0.79$) but also undershoots. Above $T_c$, the model shows strong quantitative agreement, staying within $7\%$ of the applied numerical benchmark and well inside the $1\sigma$--$2\sigma$ uncertainty bands between $160$ and $200$~MeV (for instance, matching $0.132$ against $0.132$ at $160$~MeV). At higher temperatures, it approaches the Stefan--Boltzmann limit from below, reaching $94\%$ of $1/3$ by $325$~MeV.

Right panel depicts the thermal evolution of the cumulant ratio $\chi_4^B/\chi_2^B$. We find that our ratio result is very HRG-like, e.g., $\simeq0.96$--$1.01$ between $70$ and $148$~MeV. This follows from the statistical-confinement weighting. For small $\phi,\bar\phi$, the quark contribution is dominated by the $z^3$ term\footnote{In terms of a dimensionless ratio involving mass, chemical potential, temperature, or a background/Polyakov field, the potential is often expanded as a polynomial,  $$\Omega_q \sim c_0 + c_1 z + c_2 z^2 + c_3 z^3 + \cdots,$$
here, $z^3$ is simply the cubic term in that expansion.
}, so baryon number is carried by three-quark clusters with $\chi_4^q/\chi_2^q=9$, i.e., $\chi_4^B/\chi_2^B=1$. The ratio then rises above $1$ from $T\simeq150$~MeV, reaches a narrow maximum of $3.82$ at $T=159.4$~MeV (FWHM $\simeq2$~MeV). It then drops to $0.57$ at $T=164$, $0.19$ at $T=200$, and $0.088$ at $T=325$~MeV, approaching the Stefan--Boltzmann value $0.0675$ from above. The maximum coincides with the steepest point of $\sigma_x$ and signals a near-critical enhancement, as expected for a critical endpoint at $\mu_B/T=2.08$, Section~\ref{sec:cepresult}. A peak of this height and narrowness is, to our knowledge, not observed in lattice results at $\mu=0$. There are no lattice QCD simulations available to quantify this observation. The $SU(2)$ flow test in Section~\ref{sec:frgres} shows that this behavior comes largely from the thermal-only mean-field approximation with the quark vacuum term included, which lowers the peak of the ratio by a factor of $2.3$. Adding meson fluctuations reduces it by roughly another $10\%$.

The three pseudocritical temperatures obtained from the maxima of $|\dd\sigma_x/\dd T|$, $|\dd\sigma_y/\dd T|$ and $\dd\phi/\dd T$ on a $1$--$2$~MeV grid coincide within the grid resolution, left panel of Fig.~\ref{fig:orderB},
\begin{equation}
T_\chi^{(l)}=T_\chi^{(s)}=T_\phi=159.4~\mathrm{MeV}, \label{eq:Tpcs}
\end{equation}
with FWHM bands $158.2\leq T \leq 160.6$ (light condensate), $157.5\leq T \leq 160.4$ (strange condensate), and $156.6\leq T \leq 160.6$~MeV (Polyakov loop potential). Estimates from a coarser $5$~MeV grid which is not shown here yield $158.0$, $157.6$, $157.6$~MeV, respectively. This is $1.4$~MeV above the upper edge of the quoted lattice range $155\leq T\leq 158$~MeV \cite{shah2026locating} and $1.9$ standard deviations above $156.5\pm1.5$~MeV \cite{hotqcd19}. The crossover test gives $R=0.043$ for step changes $5.94/1.26/0.25$~MeV. The crossover obtained is narrow, $\max|\dd\sigma_x/\dd T|\simeq10$--$13$, and steeper than the lattice one, which produces the fluctuation spike below. The leading-order curvature coefficient of the pseudocritical transition line in the $(T, \mu_B)$ plane is $\kappa_2^B=0.0165$, 
\begin{equation}
T_{pc}(\mu_B)=T_{pc}(0)\left[1-\kappa_2^B\left(\frac{\mu_B}{T_{pc}(0)}\right)^2\right],
\end{equation} 
which is fitted for $\mu_B\le240$~MeV yielding $T_{pc}(0)=159.1$~MeV. For the sake of completeness, the lattice curvature is expressed as \cite{bellwied15,bazavov17}
\begin{equation}
\frac{T_c(\mu_B}{T_c(\mu=0)}= 1 - \kappa_2^B \left(\frac{\mu_B}{T_c(\mu=0)}\right)^2+\lambda_4^B \left(\frac{\mu_B}{T_c(\mu=0)}\right)^4+\cdots, \nonumber
\end{equation}
where $\kappa_2^B=0.0149\pm0.0021$

\subsection{Equation of state and Susceptibilities}

\begin{figure}[htb]
\centering
\includegraphics[width=\textwidth]{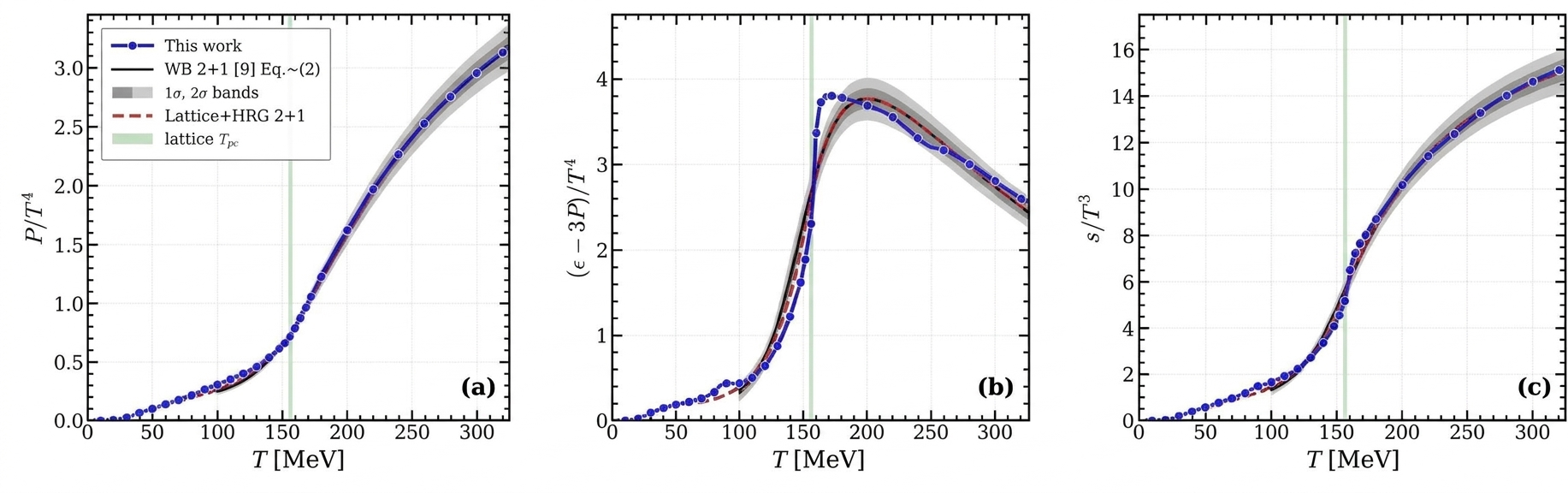}
\caption{Our calculations for the QCD equation of state at $\mu=0$ over $0\le T\le325$~MeV. Panel (a) is devoted to scaled pressure $P/T^4$, panel (b) presents scaled trace anomaly $(\epsilon-3P)/T^4$, and panel (c) shows dimensionless entropy density $s/T^3$. Black curves illustrate Wuppertal--Budapest $2+1$ parametrization, Eq.~(2) of Ref.~\cite{borsanyi14}, integrated from $T=0$ and shown for $T\ge100$~MeV. The gray bands are $1\sigma$ and $2\sigma$ ranges from the uncertainty model ($3\%$ relative, half the spread to the lattice+HRG table, and an absolute floor of $0.003$, $0.05$ and $0.05$ for $P/T^4$, $(\epsilon-3P)/T^4$ and $s/T^3$ respectively). Red dashed curves are the $2{+}1$ lattice+HRG table of Ref.~\cite{alba18}. Vertical bands mark lattice $T_{pc}=155$--$158$~MeV. 
}
\label{fig:eosB}
\end{figure}

Figure~\ref{fig:eosB} presents a comparative study for QCD EoS. Panel (a) depicts scaled pressure, where for $T\ge200$~MeV the present model reproduces the lattice results to within $1.4\%$ and lies inside the $1\sigma$ band. Over $130\le T\le325$~MeV the largest relative deviation found is $11.8\%$ at $130$~MeV. About $89\%$ ($96\%$) of the temperatures lie inside the $1\sigma$ ($2\sigma$) band. Below $130$~MeV the model lies above the lattice by up to $24\%$ in the range $100\leq T \leq 130$~MeV. This is because the pseudo-Goldstone octet gas is too large there. Compare this with the origin of the little horn that shall be discussed shortly. As expected for a crossover, the pressure has no visible structure at $T_{pc}$, i.e., the transition shows up in the derivatives. For lattice QCD scaled susceptibility,
$$\frac{\chi^B_2(T)}{T^2} = \exp\left(-\frac{h_1}{t}-\frac{h_2}{t^2}\right)\left[h_3 + f_3\frac{\tanh(f_4 t + f_5)+1}{1 + g_3 t + g_4 t^2}\right],$$
where $t=T/T_0$ with $T_0$ is a reference temperature scale which in Wuppertal--Budapest fits reads $T_0=200$~MeV and $(h_1,h_2,h_3,f_3,f_4,f_5,g_3,g_4)=
(-0.2286,0.5177,0.3333,0.002975,3.758,-2.404,-0.4408,0.06808)$.

Panel (b) is devoted to scaled trace anomaly (interaction measure). This is the most sensitive observable and carries nearly all of the $\chi^2$. The model rises more slowly than the lattice for $T\lesssim150$~MeV so that relative deviations up to $27\%$ is found at $142$~MeV. Then the model increases more steeply across $T_{pc}$ and overshoots, e.g., reaching $3.8$ at $165\leq T \leq 175$~MeV where the lattice susceptibility is $3.1$--$3.4$. The maximal absolute deviation is determined as $0.67$ at $T=163$~MeV. The maximum value that the model give, namely, $3.80$ at $T=170$~MeV, seems to match the lattice height value, $3.77$, but is $\simeq30$~MeV lower in temperature. For $T\ge200$~MeV the model lies below the lattice by at most $4.8\%$. We also estimate that over $130\leq T \leq 325$~MeV only $51\%$ ($77\%$) of the points lie within the $1\sigma$ ($2\sigma$) band. The little horn near $85\leq T \leq 90$~MeV is visible in all three panels is an artifact, namely, as $\sigma_x$ decreases below $f_{\pi}$ the tree-level pion curvature mass squared decreases from $1.9\times10^4$~MeV$^2$ in the vacuum to zero at $\sigma_x\simeq86.7$~MeV. Furthermore it would turn negative as well. The condensate is pinned near this point for $85\lesssim T \lesssim 100$~MeV, while the smoothly clamped tachyonic mass acts as a nearly massless pion gas. For the lattice QCD parametrization the peak reaches $3.77$,
\begin{eqnarray}
\frac{\epsilon-3p}{T^4}=\exp\left(-\frac{h_1}{t}-\frac{h_2}{t^2}\right)\left[h_0+f_0\frac{\tanh(f_1t+f_2)+1}{1+g_1t+g_2t^2}\right],
\end{eqnarray} 
where $t=T/200$~MeV,  $h_0=0.1396$, $h_1=-0.18$, $h_2=0.035$, $f_0=1.05$, $f_1=6.39$, $f_2=-4.72$, $g_1=-0.92$, and $g_2=0.57$.

\begin{figure}[htb]
\centering
\includegraphics[width=0.45\textwidth]{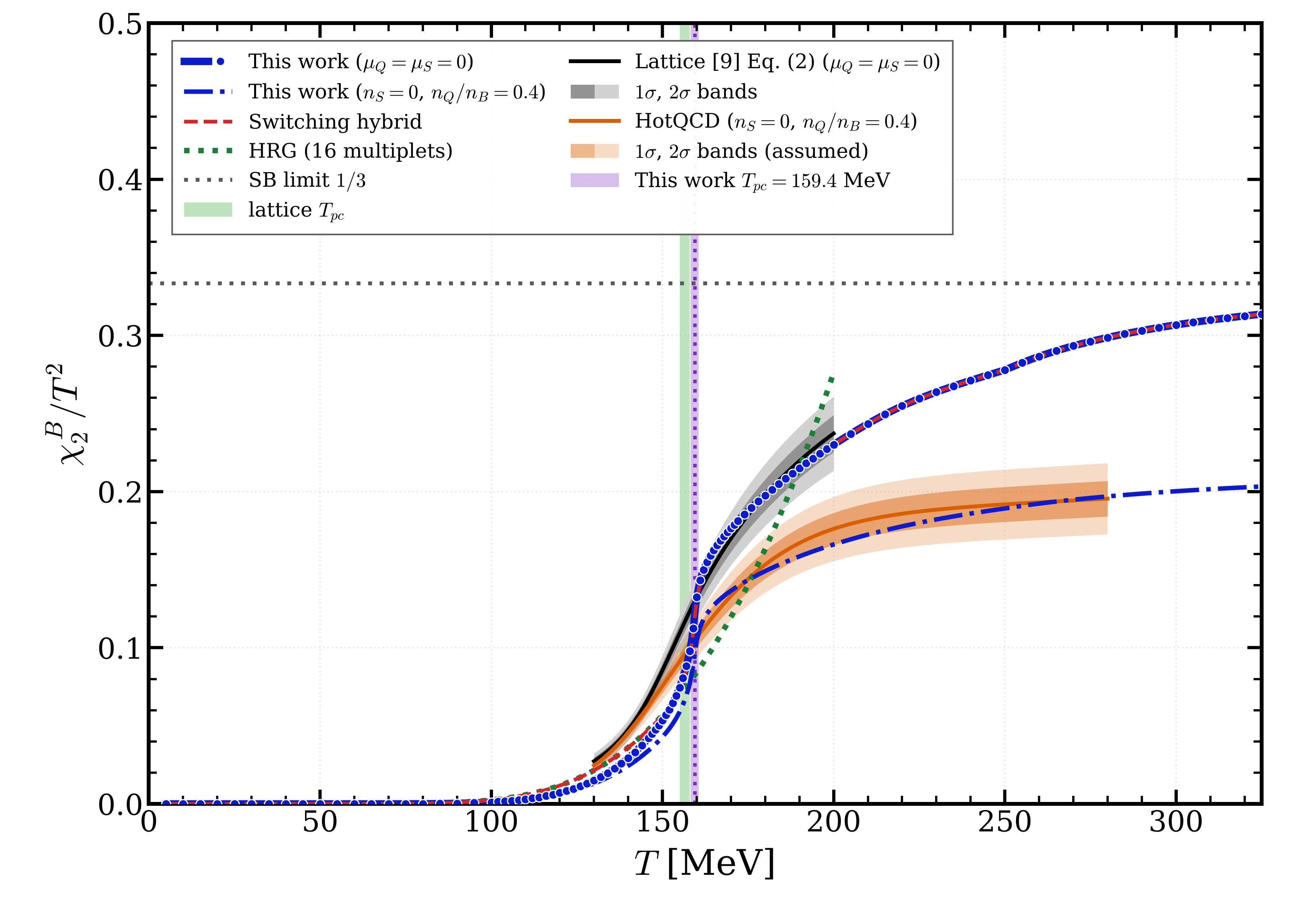}
\includegraphics[width=0.45\textwidth]{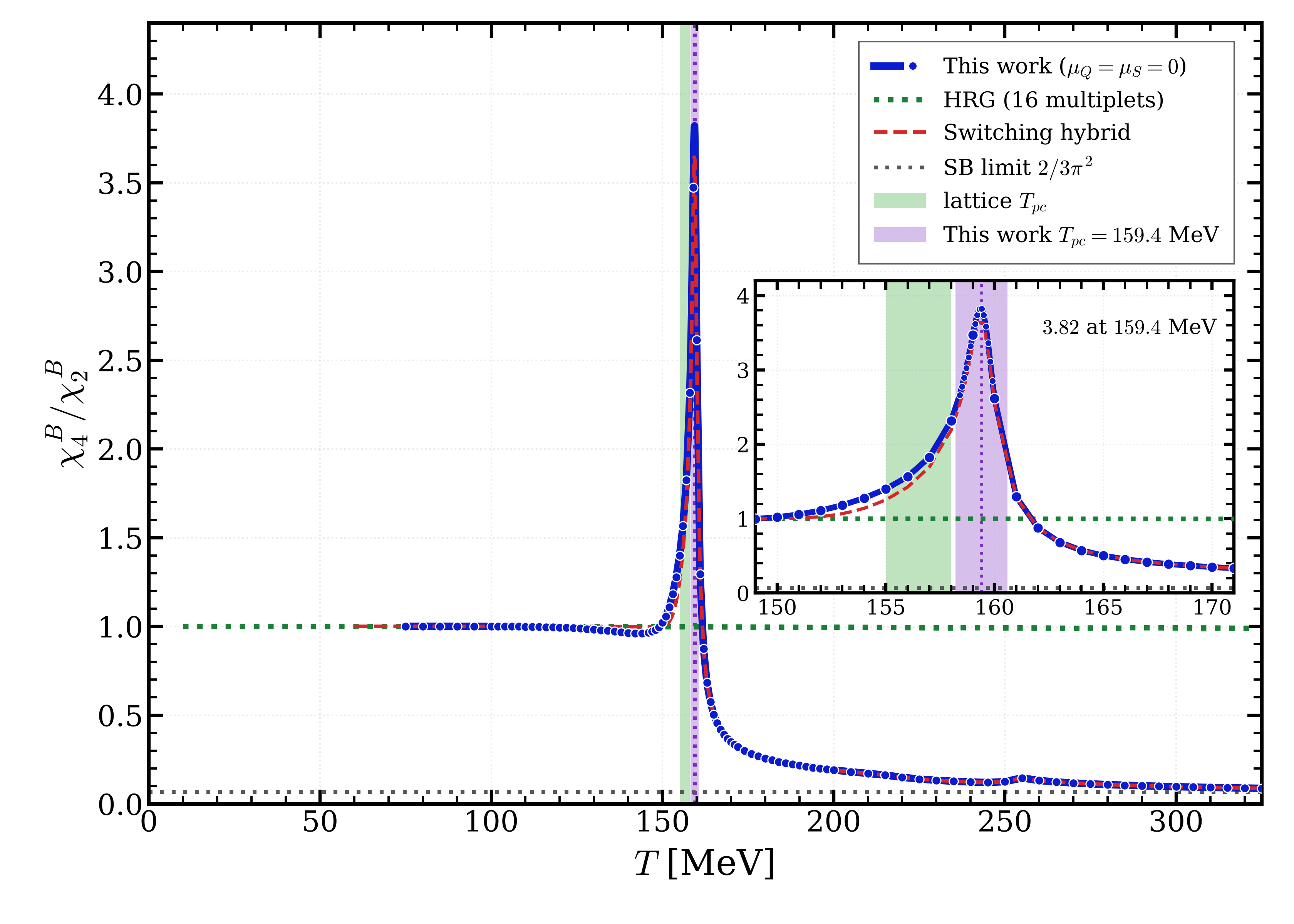}
\caption{Baryon-number susceptibility $\chi_2^B/T^2$ (left) and cumulant ratio $\chi_4^B/\chi_2^B$ (right) versus temperature $T$. Left: This work at $\mu_Q=\mu_S=0$ (solid line, circles at computed $T$) and strangeness-neutral matter ($n_S=0$, $n_Q/n_B=0.4$; dash--dotted), alongside lattice Ref.~\cite{shah2026locating} at $\mu_Q=\mu_S=0$ for $130\le T\le200$~MeV (gray $1\sigma$/$2\sigma$ bands, covariance plus $5\%$ systematic), HotQCD \cite{bazavov17} for strangeness-neutral matter over $130\le T\le280$~MeV (orange, assumed $5.8\%$/$11.7\%$ $1\sigma$/$2\sigma$ bands), 16-multiplet HRG (dotted), and HRG-to-present switching hybrid ($T_s=154$, $\Delta T=2$~MeV; dashed). Vertical bands mark lattice $T_{pc}=155$--$158$~MeV and this work's $T_{pc}=159.4$~MeV ($\sigma_x$ FWHM $158.2$--$160.6$~MeV); gray dotted line marks the $\mu_Q=\mu_S=0$ Stefan--Boltzmann limit ($1/3$). Right: Cumulant ratio $\chi_4^B/\chi_2^B$ at $\mu_Q=\mu_S=0$ from this work (thick blue line with circles; inset shows $149\le T\le171$~MeV with $0.1$~MeV points near peak), dropping $\chi_2^B < 2\times10^{-5}$ ($T \lesssim 75$~MeV) due to finite-difference noise. Shown with 16-multiplet HRG to $200$~MeV (green dotted), hybrid (red dashed), Stefan--Boltzmann limit $2/3\pi^2=0.0675$ (gray dotted), lattice $T_{pc}$ (green band), and this work's $T_{pc}=159.4$~MeV (purple line/band).
}
\label{fig:hrg}
\end{figure}

Turning to detailed description of susceptibilities, we compare our results against the $\chi_2^B$ lattice parametrization of Ref.~\cite{shah2026locating}, using $3000$ draws from its published covariance matrix (with a $5\%$ floor) to define the $1\sigma$ uncertainty band. Within $160\le T\le 200$~MeV, the agreement is very close. We obtain $\chi^2=4.1$ across $11$ points and a maximum deviation of $6.8\%$, matching $0.1323$ against $0.1323$ at $T=160$~MeV and $0.2301$ against $0.2372$ at $T=200$~MeV. Below the crossover, $120\le T<160$~MeV, however, the model undershoots the lattice results by a factor of $0.44$--$0.71$, yielding $\chi^2=316$ over $10$ points. Finally, at high temperatures, $\chi_2^B/T^2$ reaches $0.305$ at $T=295$~MeV, approaching the $1/3$ SB-limit from below.

As in middle panel of Fig.~\ref{fig:orderB}, left panel of Fig.~\ref{fig:hrg} does not only display 16-multiplet HRG, Section~\ref{sec:num}, but also strangeness neutral $\mu_Q=\mu_S=0$ and switching hybrid $\chi_2^{\rm hyb}=(1-s)\chi_2^{\rm HRG}+s\chi_2^{\rm B}$, where $s=[1+e^{-(T-T_s)/\Delta T}]^{-1}$. We find that for strangeness-neutral calculations below $T_c$, our model undershoots the lattice parametrization (ratio $0.57$--$0.71$ for the temperature range $130\leq T\leq 156$~MeV). While 16-multiplet HRG is closer ($0.63$--$0.79$) but it still overestimates. As expected from the missing resonances, HRG overshoots above $191.5$~MeV (by $16\%$ at $T=200$~MeV). Thus, HRG is only meaningful below $T_c$. On the other hand, the switching hybrid model seems to track the HRG below $T_c$ nearly perfectly and matches the results of our model above it. Across $120\leq T\leq 200$~MeV (21 points), it reduces the $\chi^2$ against the lattice parametrization from $320$ to $176$ which is driven by a drop from $316$ to $172$ below $T=160$~MeV, while staying at $4.1$ above. Picking $T_c=154$~MeV which is roughly $5\text{ MeV}$ below the model $T_{pc}$, came from a coarse grid simulation, so part of the gain is simply fitting. Because the hybrid blends the two regimes together afterward as a post-processing step. It is not guaranteed to stay thermodynamically consistent.  Above $T_c$ for strangeness neutrality $\mu_Q=\mu_S=0$, the model stays within $6.8\%$ of lattice results across $160\leq T\leq 200$~MeV ($\chi^2=4.1$ over 11 points) and climbs toward $1/3$ from below, reaching $0.314$ at $T=325\text{ MeV}$. Across temperatures from $160$ to $280\text{ MeV}$, our strangeness-neutral calculations match the HotQCD parametrization within $5.8\%$. The largest gap appears at $T=164\text{ MeV}$, and overall goodness of fit gives $\chi^2=44.9$ across $121$ data points. In practical terms, $98\%$ of the points fall inside the expected band, and every single point stays within the $1\sigma$ and $2\sigma$ bounds. For $200\le T\le280$~MeV the maximum deviation is $5.7\%$ and $\chi^2=24.3$ for 81 points. The ratio model/HotQCD is found $0.98$, $0.97$, $0.94$, $0.99$ and $1.01$ at $T=160$, $180$, $200$, $250$ and $280$~MeV, respectively. Below $T_c$ the model is again too low, namely, ratios $0.53$, $0.53$, $0.57$, respectively, are obtained at $T=130$, $140$, $150$~MeV. It seems that the effect of strangeness neutrality includes suppression of the baryon susceptibility. The ratio $2P_2/\chi_2^B(\mu_Q=\mu_S=0)$ is evaluated as $0.79$, $0.76$, $0.72$ in the model, at $T=160$, $180$, $200$~MeV, respectively, against the values $0.81$, $0.77$, and $0.74$ for HotQCD over the lattice parametrization, i.e., within $3\%$ for $T\ge160$~MeV. At $T=140$~MeV, it increases from $0.83$ against $0.97$. In our model the ratio reaches $0.66$ at $T=280$~MeV. In this regard, we emphasize that the HotQCD band is assumed, while HotQCD parametrization is used only within its stated range, namely $130\leq T\leq 280$~MeV,  because its rational form is not valid at larger $T$. 

The right panel presents the cumulant ratio $\chi_4^B/\chi_2^B$ as a function of temperature. In hadronic regime  for $80\le T\le148$ MeV, we find that the ratio is about $0.96-1.00$, i.e., reproduces well the HRG value which varies between 1.0 at low $T$ and $\sim 0.993$ at $T=200$~MeV. In our model, this value comes from three-quark clusters carrying baryon number. The agreement is within $4\%$, especially below the transition.
During the transition, the ratio leaves the HRG value near $T=150$~MeV and has a single maximum of height $3.82$ at $T=159.4$~MeV. The full width at half maximum is about $3.2$~MeV within $157.2\leq T\leq 160.4$~MeV. This result coincides well with $T_{pc}$ and with the steepest rise of $\chi_2^B$, left panel. We find that the height is stable against the finite-difference step but depends on the temperature sampling so that its height reaches $2.6$ on a $2$~MeV grid, $3.5$ on $1$~MeV, and $3.82$ on $0.1$~MeV. This is why the inset is displayed. We also notice that the hybrid calculations cannot remove the peak so $2.57$ against $2.62$ are found at $T=160$~MeV. At high $T$, the ratio falls monotonically; from $0.35$ at $T=170$~MeV to $0.19$ at $=200$~MeV, to 0.097 at $T=300$~MeV and to 0.088 at $T=325$~MeV. The last value is just 1.3 times higher than the Stefan--Boltzmann limit ($2/3\pi^2=0.0675$ dotted line). There is a weak bump near $T=255$~MeV. It seems that this result coincides well with the secondary structure in the strange condensate (left panel of Fig.~\ref{fig:orderB}).

We assume that the peak comes primarily from omitting the vacuum term in our mean-field calculation. Our $SU(2)$ test backs this up so that including the vacuum contribution suppresses the peak by a factor of $2.3$, whereas accounting for fluctuations shifts it by only around $10\%$ more. This observation points to a clear hierarchy in what shapes the signal. The vacuum term acts as a strong damping mechanism likely by stiffening the effective dynamics or changing the curvature of the thermodynamic potential across the transition. Thus leaving it out allows the peak to stand out prominently. Fluctuations on the other hand refine the result at the roughly $10\%$ level, but they are secondary to the choice of whether to include the vacuum term. In short, we conclude that the prominent peak is largely a feature of a vacuum-term-free mean-field setup rather than a fluctuation-driven effect.

Thus, we conclude that below the transition the SPLSM reproduces the HRG results of the cumulant ratio. It shows one narrow maximum of height $3.82$ at $T_{pc}$. We consider this a mean-field feature not a QCD prediction. A hadronic patch does not seem to remove it. Above the transition the ratio decreases toward the Stefan--Boltzmann limit. To establish these findings more firmly, we need a quantitative comparison with lattice QCD results, examining magnitudes, temperature dependence, and uncertainties throughout the crossover region.

With these ingredients in place, we can now map out the phase diagram and pinpoint the critical endpoint (see Section~\ref{sec:cepresult}).

\subsection{Phase diagram and critical endpoint}\label{sec:cepresult}

\begin{figure}[htb]
\centering
\includegraphics[width=0.78\textwidth]{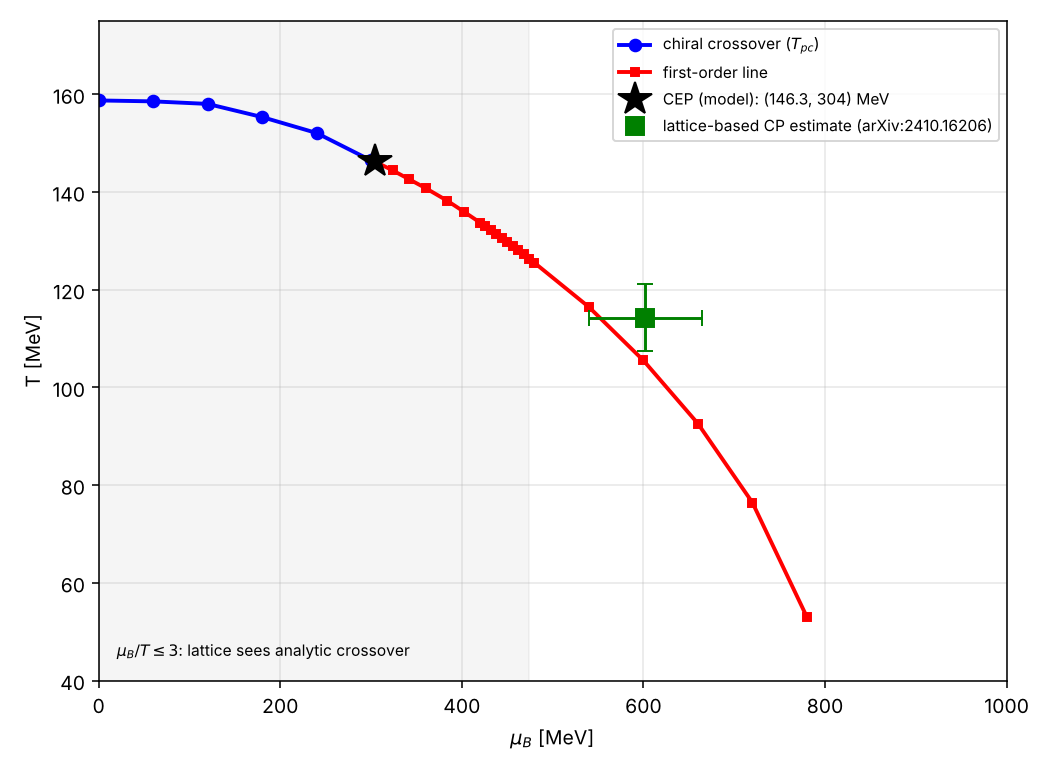}
\caption{Phase diagram drawn from applied computational benchmark, where $\mu_B=3\mu_q$. Blue curved stands for chiral crossover.  The red curve shows first-order boundary determined at $\Omega_{\rm b}=\Omega_{\rm r}$, i.e., coexistence method. Star positions CEP $(146.3,\,304.1)$~MeV. Big square with errorbars refers to the lattice-based critical-point estimate at $(114.3\pm6.9,\ 602.1\pm62.1)$~MeV of Ref.~\cite{shah2026locating}. Shaded region marks the entire domain $\mu_B/T\le3$, where lattice QCD simulations presumably expect analytic crossover. The symbols associated with both curves mark $\mu_B$-values at which the numerical analyses were performed. }
\label{fig:phaseB}
\end{figure}

\begin{figure}[htb]
\centering
\includegraphics[width=\textwidth]{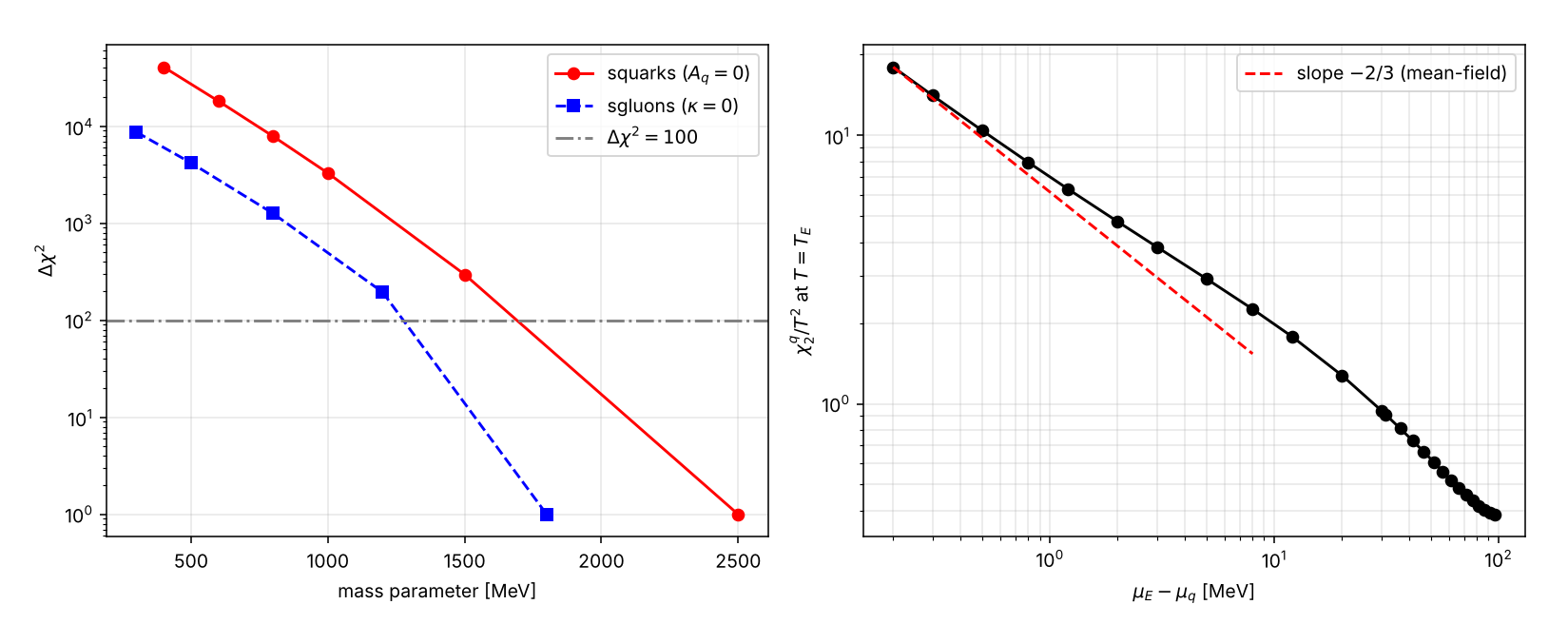}
\caption{Left panel: Change in $\Delta\chi^2$ across temperatures from $130$ to $280$~MeV relative to our computational benchmark, evaluated with re-profiled $T_0$ for a free-gas squark sector at $A_q=0$ (solid curve) and a sgluon sector at $\kappa=0$ (dashed curve). Right panel: Rescaled quark susceptibility $\chi_2^q/T^2$ along the critical isotherm, compared against the mean-field scaling slope of $-2/3$. }
\label{fig:sens}
\end{figure}

Starting from four independent initial guesses, the numerical solver consistently converges to a unique critical endpoint (CEP) at
\begin{equation}
T_E = 146.253~\mathrm{MeV}, \qquad \mu_q^E = 101.379~\mathrm{MeV}, \qquad \mu_B^E = 304.14~\mathrm{MeV}, \quad \text{and} \quad \mu_B^E/T_E = 2.08, 
\label{eq:cepB}
\end{equation}
satisfying the criticality conditions in Eqs.~\eqref{eq:cep1}--\eqref{eq:cep3} to a residual of $10^{-13}$. At this point, the soft mode is dominated by the non-strange condensate $\sigma_x$, with eigenvector components $(0.982, 0.190, -0.002, -0.002)$. Three independent consistency checks confirm that this endpoint is genuine rather than a numerical artifact:
\begin{enumerate}
\item Mean-field scaling: Fitting the square-root scaling law of Eq.~\eqref{eq:mfscaling} to the order-parameter jumps across $\mu_q = 108,\dots,134~\mathrm{MeV}$, with $\Delta\sigma_x$ grows from $13.9$ to $34.1~\mathrm{MeV}$, in that order, extrapolates to a critical chemical potential of $\mu_q = 102.6~\mathrm{MeV}$, agreeing with $\mu_q^E$ to within $1.3~\mathrm{MeV}$.
\item Crossover continuity: The pseudocritical crossover line approaches the endpoint smoothly from lower chemical potentials, reaching $T_{pc}=146.5~\mathrm{MeV}$ at $\mu_B=300~\mathrm{MeV}$.
\item Critical isotherm divergence: Along the critical isotherm, $T=T_E$, the quark number susceptibility follows the power law $\chi_2^q \propto \vert{}\mu_E - \mu\vert{}^{-p}$. The effective exponent increases from $p=0.56$ for $\delta\mu < 8~\mathrm{MeV}$ to $p \simeq 0.60$ for $\delta\mu \lesssim 0.5~\mathrm{MeV}$, steadily approaching the mean-field value of $2/3$ as $\delta\mu \to 0$ (see Fig.~\ref{fig:sens}).
\end{enumerate}
Beyond the CEP, the first-order phase transition strengthens rapidly. The condensate discontinuity $\Delta\sigma_x$ widens from $37.5~\mathrm{MeV}$ at $\mu_q=140~\mathrm{MeV}$ to $47.2~\mathrm{MeV}$ at $\mu_q=160~\mathrm{MeV}$. As this first-order boundary extends toward higher baryon densities, the transition temperature drops sharply so sitting at $T=125.5~\mathrm{MeV}$ for $\mu_B=480~\mathrm{MeV}$, dipping to $105.5~\mathrm{MeV}$ at $\mu_B=600~\mathrm{MeV}$, and falling to $53.0~\mathrm{MeV}$ by $\mu_B=780~\mathrm{MeV}$.

Taken together, our SPLSM provides strong evidence for a single, well-defined critical endpoint that holds up under numerical checks. Converging to the exact same coordinates from four independent initial guesses confirms that the solution is a genuine, isolated critical point rather than a local artifact or starting-point-dependent root. A residual of $10^{-13}$ obviously demonstrates high numerical stability in solving the criticality conditions. In this benchmark, the QCD phase transition switches from a smooth crossover to a first-order transition at $T_E \approx 146.3~\mathrm{MeV}$ and $\mu_B^E \approx 304.1~\mathrm{MeV}$, i.e., $\mu_B^E/T_E=2.08$. We assert that these eigenvector components seem to point to light-quark chiral symmetry restoration as the primary driver of the transition. Because the critical soft mode aligns so strongly with the non-strange condensate, $\sigma_x$ ($0.982$) picking up only a modest strange-condensate contribution ($0.190$) and a negligible Polyakov-loop component ($-0.002$), deconfinement degrees of freedom likely play a largely passive role near the CEP. Across the phase boundary, there is smooth continuity that the pseudocritical crossover line, $T_{pc}=146.5~\mathrm{MeV}$ at $\mu_B=300~\mathrm{MeV}$,  connects seamlessly into the CEP at $T_E= 146.253~\mathrm{MeV}$ and $\mu_B^E= 304.14~\mathrm{MeV}$), confirming that the crossover and first-order lines form a single continuous phase boundary. This allows us to emphasize that both signatures likely point to standard mean-field critical scaling, matching what this truncation should yield: the condensate jump $\Delta\sigma_x$ above $\mu_q^E$ exhibits square-root scaling that extrapolates within $1.3~\mathrm{MeV}$ of the direct solver result, and the quark number susceptibility $\chi_2^q$ along the critical isotherm diverges with an effective exponent $p \to 2/3$ as $\delta\mu \to 0$. Also, we find rapid strengthening and steepening of the first-order transition at high density. Once past the CEP, the transition becomes increasingly discontinuous so that the jump $\Delta\sigma_x$ grows substantially over a relatively narrow chemical-potential window, $37.5 \to 47.2~\mathrm{MeV}$ between $\mu_q=140$ and $160~\mathrm{MeV}$. Simultaneously, the first-order coexistence curve bends sharply downward in the $(T, \mu_B)$ plane. It drops by more than $70~\mathrm{MeV}$ in temperature between $\mu_B = 480~\mathrm{MeV}$ and $\mu_B=780~\mathrm{MeV}$. This indicates that dense cold quark matter likely transitions at much lower temperatures.

In this context, it is essential to assess how sensitive our results are to the squark and sgluon sectors. We examine this question in detail in Section~\ref{sec:susyres}.

\subsection{Sensitivity to squarks and sgluons}\label{sec:susyres}

\begin{table}[htb]
\centering
\begin{tabular}{| c | c | c | c | c | c |}
\hline \hline
$m_0$ [MeV] & $\chi^2$ & $\Delta\chi^2$ & $m_O$ [MeV] & $\chi^2$ & $\Delta\chi^2$ \\
\hline \hline
400 & 41025 & 40769 & 300 & 8971 & 8715 \\
600 & 18442 & 18186 & 500 & 3897$^\dagger$ & 3641 \\
800 & 6682$^\dagger$ & 6426 & 800 & 1519 & 1263 \\
1000 & 3409$^\dagger$ & 3153 & 1200 & 451 & 195 \\
1500 & 553 & 297 & 1800 & 253 & $-3$ \\
2500 & 252 & $-4$ & 500 ($\kappa=20$) & 4407 & 4151 \\
\hline
\end{tabular}
\caption{EoS sensitivity to a colored scalar sector relative to the present work ($\chi^2_{\rm ref}=256$, $N=76$). $T_0$ is re-profiled over $\pm8$~MeV and also over $+8$ to $+40$~MeV for entries marked $^\dagger$ to define the best fit on the edge. Low-mass entries without $^\dagger$ represent upper bounds.}
\label{tab:sens}
\end{table}

Carrying $36$ dof compared to just $16$ for sgluons, the $Z_3$-neutral squark sector is far more constraining. In the heavy-mass limit, both sectors decouple cleanly, leaving only minor shifts of $\Delta\chi^2=-4$ at $m_0=2500~\mathrm{MeV}$ and $\Delta\chi^2=-3$ at $m_O=1800~\mathrm{MeV}$. Meanwhile, turning on the portal coupling $\kappa=20$ changes the sgluon contribution by roughly $1\%$ even down at $m_O=500~\mathrm{MeV}$. The reason behind this effect is the thermal weight of extra dof, but not the chiral coupling. In this regard, we note that the critical endpoint is almost insensitive to sgluonos and moves only for squarks lighter than about $600$~MeV. We have tracked it by continuation from the decoupled limit with $T_0$ fixed, so that $(T_E,\mu_B^E)=(146.2,304.2)$ for $m_0\ge1500$, $(145.5,304.7)$ at $m_O=1000$~MeV, $(144.3,303.6)$ at $m_O=800$~MeV, and $(141.8,294.5)$ at $m_O=600$~MeV. The numerical results for $T_E$~[MeV], $\mu_B^E$~[MeV], and $\mu_B^E/T_E$ with varying masses of squarks and sgluons are listed in the following table:
\begin{center}
\begin{tabular}{| c | c | c | c | c |}
\hline\hline 
 & mass [MeV] & $T_E$ [MeV] & $\mu_B^E$ [MeV] & $\mu_B^E/T_E$ \\ \hline \hline
decoupled & $\ge1500$ & 146.2 & 304.2 & 2.08 \\ \hline \hline
squarks & 800 & 144.3 & 303.6 & 2.10 \\
squarks & 600 & 141.8 & 294.5 & 2.08 \\
squarks & 500 & 140.0 & 279.2 & 1.99 \\
squarks & 450 & 139.1 & 265.2 & 1.91 \\
squarks & 400 & 138.4 & 243.7 & 1.76 \\ \hline \hline
sgluons & 1200 & 146.2 & 304.0 & 2.08 \\
sgluons & 800 & 145.8 & 303.3 & 2.08 \\
sgluons & 500 & 144.2 & 301.8 & 2.09 \\
sgluons & 300 & 141.2 & 303.5 & 2.15 \\
\hline
\end{tabular}
\end{center}
We find that the light squarks move the endpoint to lower temperature and lower baryon chemical potential so that at $m_0=400$~MeV it is moved by $60$~MeV in $\mu_B$ and $8$~MeV in $T_E$. Thereby, it reduces $\mu_B/T$ from $2.08$ to $1.76$, i.e., the light squarks push endpoint further into the region excluded by lattice extrapolations. Sgluons shift $T_E$ by up to $5$~MeV and leave $\mu_B^E$ unchanged to within $2.5$~MeV down to $m_O=300$~MeV. The $\mu=0$ transition remains an analytic crossover for all squark masses tested down to $400$~MeV with step-scaling $R=0.045$--$0.048$ for $m_0=600$--$400$~MeV, respectively. Thus the endpoint does not run into $\mu=0$. The continuation at $m_0=450$~MeV requires a starting point interpolated between the $500$ and $400$~MeV solutions.

Section~\ref{sec:frgres} details the fluctuation test in the $SU(2)$ sector. Testing fluctuations in this reduced setting lets us isolate the role of specific model ingredients, such as vacuum terms and crossover behavior, and benchmark the numerical machinery against a cleaner, lower-rank gauge group.

\subsection{Fluctuation test in the $SU(2)$ sector}
\label{sec:frgres}

\begin{figure}[htb]
\centering
\includegraphics[width=\textwidth]{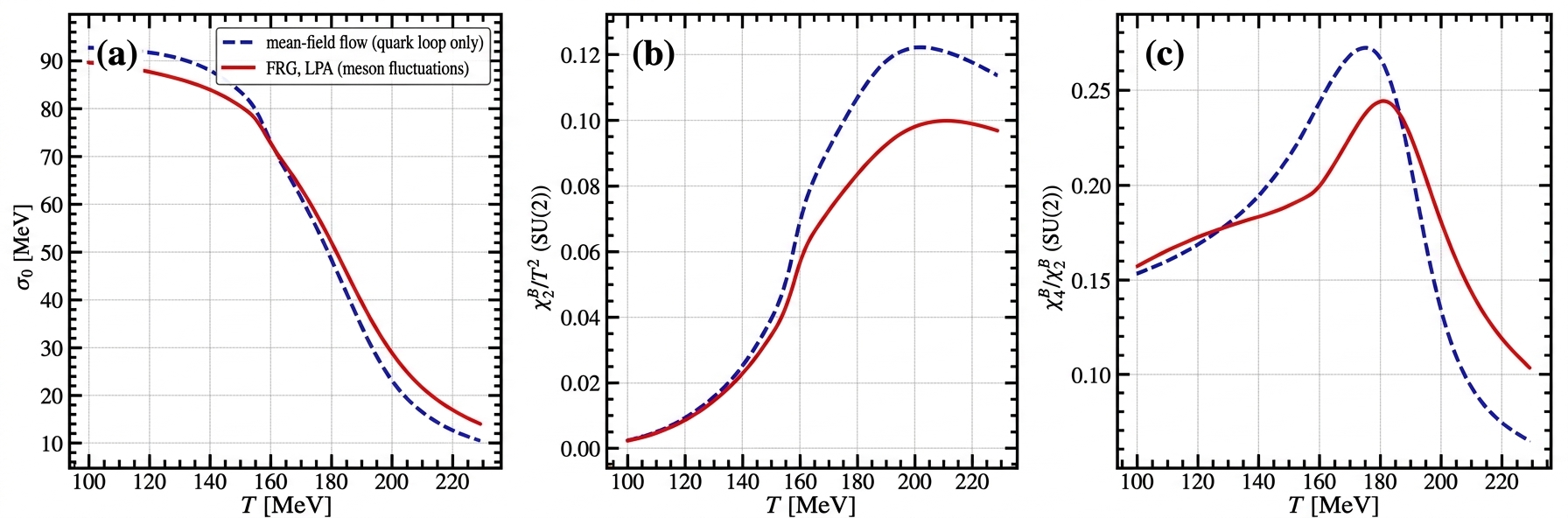}
\caption{Thermal evolution in the $SU(2)$ quark-meson model ($m_\sigma=500$~MeV) comparing the mean-field flow with the quark loop (dashed curves) against the FRG in the LPA, evaluated under the same UV initial conditions and Polyakov background. Left panel shows $\sigma_0(T)$, while middle panel presents $\chi_2^B/T^2$ and right panel is devoted to $\chi_4^B/\chi_2^B$.}
\label{fig:frg}
\end{figure}

At $\mu=0$, we find that with the UV parameters of Section~\ref{sec:frgmeth}, the half-restoration temperature $T(\sigma=f_\pi/2)$ reads $182.1$~MeV in the mean-field flow and $184.9$~MeV in the FRG, i.e., there is a shift of $+2.7$~MeV. The maximum slope $|\dd\sigma/\dd T|$ is determined as $1.446$ at $T=158.6$~MeV in the mean-field flow. This is a feature which is driven by the steep rise of the prescribed $\phi(T)$ at $T\simeq158$~MeV. A maximum slope of $1.279$ is determined at $T=184.4$~MeV in the FRG, where the background-induced feature is likely washed out. The minimal $\sigma$ mass is evaluated as $283$ and $245$~MeV. The susceptibilities and MF flow versus FRG are summarized in the following table:
\begin{center}
\begin{tabular}{| c | c | c | c | c |}
\hline \hline
$T$ [MeV] & 150 & 170 & 184 & 200 \\
\hline \hline
$\chi_2^B/T^2$, MF flow & 0.0403 & 0.0916 & 0.1122 & 0.1221 \\
$\chi_2^B/T^2$, FRG & 0.0353 & 0.0729 & 0.0880 & 0.0982 \\ \hline \hline
$\chi_4^B/\chi_2^B$, MF flow & 0.215 & 0.267 & 0.251 & 0.132 \\
$\chi_4^B/\chi_2^B$, FRG & 0.190 & 0.225 & 0.243 & 0.180 \\
\hline
\end{tabular}
\end{center}
We find that meson fluctuations reduce $\chi_2^B$ by $12$--$22\%$ and drop the maximal of $\chi_4^B/\chi_2^B$ from $0.27$ to $0.24$. The thermal-only $SU(2)$ mean field, where same is valid for $g$, $m_\sigma$ and background and no vacuum term is considered, seems to give a maximum of $0.63$ at $T=163$~MeV. Furthermore, the vacuum loop reduces the peak by a factor $2.3$ and meson fluctuations by a further $\sim10\%$.

Now for critical endpoint, three treatments with identical $m_\sigma$, $g$, and background are compared with each other in the following table:\\
\begin{tabular}{| c | c | c |}
\hline\hline
Treatment & $\mu_q^E$ [MeV] (indicator) & $T_E$ [MeV], $\mu_B^E$ [MeV] \\
\hline \hline
Thermal-only mean field (no vacuum) & $167\pm3$ (step scaling: $R=0.077$ at 165, $0.158$ at 170) & $\simeq148$, $501\pm9$ \\
Mean-field flow (vacuum loop included) & $\simeq286$ (bracket 285--290; extrapolation of $(m_{\sigma,\min}^2)^{3/2}$) & $\simeq106$, $\simeq859$ \\
FRG, LPA (up to 5th-order Taylor) & $\simeq253\pm3$ (onset of expansion failures) & $\simeq121$, $\simeq759$ \\
\hline
\end{tabular}
It is obvious that in the mean-field flow, breakdowns of the Taylor expansion occur only once the transition turns first order, i.e., at $\mu_q\ge 290$~MeV, matching the extrapolated critical endpoint. In the FRG, by contrast, this instability band opens up earlier --- growing from zero width at $\mu_q\simeq 253$~MeV to widths of $2.0, 7.5, 11.5, 16.5,$ and $21.5$~MeV across $\mu_q=255,\dots,275$~MeV, respectively. Because the chiral condensate $\sigma$ jumps discontinuously across this region, we use its onset as a practical indicator of the endpoint. Note, however, that this criterion serves only as a proxy, since an expansion around a single minimum cannot resolve the full phase coexistence region. Finally, we do not explicitly analyze the numerical breakdown region at $\mu_q\ge 320$~MeV. Within this limitation, three conclusions can be drawn: 
\begin{itemize}
\item meson fluctuations do not remove the endpoint in LPA but shift it by $\sim12\%$ in $\mu_q$ to smaller $\mu$ than the mean-field flow, i.e., not to a larger $\mu$ as often assumed,
\item the vacuum quark loop shifts the endpoint by $\sim360$~MeV in $\mu_B$, i.e., far more than fluctuations do, and  
\item hence the low-$\mu_B$ endpoint of the applied computational benchmark, $\mu_B/T\simeq 2.1$, and its $\chi_4$ spike are mainly artifacts of the thermal-only (without vacuum term) mean-field treatment, which seems to be consistent with results on  Ref.~\cite{skokov10a}. 
\end{itemize}
In the $SU(2)$ flows, the endpoint has $\mu_B/T\gtrsim6$. Apparently, a full $2{+}1$-flavour FRG is needed before drawing the same conclusion for the $SU(3)$ endpoint quantitatively.

To round out the numerical analysis, Section~\ref{sec:FlvrsX} explores how flavor extensions could affect the model. Varying the flavor content, such as moving between two- and three-flavor sectors or adjusting quark mass hierarchies, helps test how sensitive the transition structure, condensates, and susceptibilities are to the underlying flavor dynamics.

\subsection{Flavour extensions}
\label{sec:FlvrsX}

Turning to flavor extensions, we begin with the two-flavor limit. While $SU(2)$ enters our analysis through the quark--meson fluctuation test and the holomorphic superpotential construction of Eq.~\eqref{eq:Wsu2}, we did not perform a full numerical run of the $SU(2)$ PLSM coupled to squarks and sgluons. In the opposite direction, extending the flavor sector to $SU(4)$ introduces the charm quark. To gauge its thermal impact, we perform a controlled benchmark by adding a Polyakov-loop-weighted free charm quark with constituent mass $m_c$ and comparing its contributions to the normalized pressure $\Delta(P/T^4)$ and trace anomaly $\Delta I/T^4 \equiv \Delta[(\epsilon - 3p)/T^4]$ against the difference between $(2+1+1)$- and $(2+1)$-flavor lattice QCD results:
\begin{center}
\footnotesize
\begin{tabular}{|c | c | c | c | c | c | c | c | c |} \hline \hline 
 & \multicolumn{2}{c|}{Lattice} & \multicolumn{2}{c|}{$m_c=1100$~MeV} & \multicolumn{2}{c|}{$m_c=1275$~MeV} & \multicolumn{2}{c|}{$m_c=1500$~MeV} \\
$T$ [MeV]    & $\Delta P$ & $\Delta I$ & $\Delta P$ & $\Delta I$ & $\Delta P$ & $\Delta I$ & $\Delta P$ & $\Delta I$ \\ \hline \hline 
200 & 0.0136 & 0.1325 & 0.0373 & 0.2041 & 0.0186 & 0.1177 & 0.0074 & 0.0551 \\
240 & 0.0418 & 0.1893 & 0.0886 & 0.3630 & 0.0509 & 0.2440 & 0.0243 & 0.1386 \\
280 & 0.0798 & 0.3150 & 0.1560 & 0.5029 & 0.0990 & 0.3773 & 0.0538 & 0.2461 \\ \hline
\end{tabular}
\end{center}
Interpolating across the tabulated masses indicates that a constituent charm mass in the range $m_c \simeq 1.3$--$1.4$~GeV reproduces the lattice charm contribution to within $\sim 20\%$ across the $200\leq T\leq 280$~MeV window, providing a useful first-order estimate between the discrete benchmark points. Several limitations apply to this comparison. First, this calculation serves as a frozen-condensate check rather than a full treatment of $SU(4)$ chiral dynamics. Second, properly capturing the thermal fluctuations of charmed mesons, such as the $D$, $D_s$, and $\eta_c$ states, requires a loop-corrected $SU(4)$ formulation. Finally, the supersymmetric sextic potential structure introduced in Section~\ref{sec:holo} has not yet been implemented in this four-flavor extension.

\section{Discussion and Conclusions}
\label{sec:disc}

We start our discussion with the physical interpretation. The Polyakov-loop coupling gives the model three distinct classes of baryon-number carriers, namely, (i) three-quark clusters below $T_c$, i.e., HRG-like $\chi_4/\chi_2=1$, (ii) single quarks above $T_c$, and, (iii) in the supersymmetric extension, $Z_3$-breaking squarks and $Z_3$-neutral sgluons. Because the sgluon is adjoint it acts as additional glue, the squark acts as a bosonic quark partner with baryon number $1/3$. Thus the EoS is sensitive to both constituents only through their thermal weights. The chiral order parameters respond through $M_{f,\pm}(\sigma_{x,y})$ and the portal coupling parameter $\kappa$. The response of the critical endpoint is small because it sits at $T\simeq145$~MeV where the TeV-scale states are likely Boltzmann suppressed.

To answer the question whether squark or sgluon effects in this model can mimic critical dynamics or confinement effects. we briefly answer with "no". A light colored scalar enters as extra thermal dof and thereby raises $P/T^4$ and the trace anomaly $I/T^4$ in the transition region. This is not degenerate with the critical-endpoint structure, which is almost unmoved, or with the Polyakov-loop calibration, where the compensating $T_0$ increase drives the transition toward first order. The lattice QCD trace anomaly is therefore a clean constraint on light colored scalars, especially in this model.

As with any effective framework, our approach entails several approximations that define its scope. The primary limitations are the free-gas treatment of the supersymmetric sector, the restriction of our FRG analysis to two flavors rather than a full $(2+1)$-flavor flow, the residual sensitivity of $T_c$ to the free sigma mass parameter $m_\sigma$, and the lack of enough tabulated lattice QCD fluctuation benchmarks across the full parameter space. Moreover, our conclusion that the low-$\mu_B$ critical endpoint arises as a vacuum-term artifact is established within the $SU(2)$ sector. It verifies whether the same mechanism persists in the $SU(3)$ case will require a dedicated refit of the corresponding strange-sector vacuum terms.

We can now draw concrete conclusions. We start with the underlying theory. Holomorphy fixes the meson potential so that the F-term of $c_W\det\hat\Phi$ gives $\lambda_1=-\lambda_2=|c_W|^2/2$ for $N_f=3$, holomorphic Mexican hat with a singlet for $N_f=2$ and a sextic term for $N_f=4$. The $m^2$ of the chiral model as well as anomaly and source terms are soft. Squarks with F-term masses $m_0^2+m_f^2\pm A_qm_f$ and sgluons which are complex adjoint scalars were coupled to the Polyakov loop. Their thermodynamics is derived from the path integral with a Matsubara Lemma and a recursion for susceptibilities to fourth order.
The applied computational benchmark results in lattice-calibrated analytic crossover at $T_{pc}=159.4$~MeV, EoS where $\chi^2/N=3.4$ especially in 130--280~MeV, and $\chi_2^B$ above $T_c$ with $\le7\%$ accuracy. On the other hand $\chi_4^B/\chi_2^B$ ratio is HRG-like below $T_c$ and approaches the Stefan--Boltzmann limit above $T_c$. We locate the critical endpoint at $(T_E, \mu_B^E)=(146.3,\,304)~\mathrm{MeV}$, corresponding to $\mu_B/T \simeq 2.08$, which sits in tension with current lattice QCD exclusion bounds. A targeted $SU(2)$ FRG benchmark traces this discrepancy primarily to the omission of the fermionic vacuum term, which shifts the endpoint by $\sim 360~\mathrm{MeV}$, whereas bosonic meson fluctuations affect the critical quark chemical potential $\mu_q$ at only the $\sim 12\%$ level.

Regarding the colored scalars, we recall that within a free-gas model the EoS fit is unchanged for $m_0\ge2.5$~TeV and $m_O\ge1.8$~TeV and degraded below $1.5$ and $1.2$~TeV. The endpoint is nearly insensitive. Note that these are sensitivity results not concrete collider limits. An HRG patch seems to improve $\chi_2^B$ below $T_c$ ($320\to176$) but not the pressure. A frozen charm quark with $m_c\simeq1.3$--$1.4$~GeV matches well the lattice charm contribution to $\sim20\%$.

As for future research directions, we highlight that $2+1$-flavour FRG including the Polyakov-loop dynamics is essential. The loop-corrected vacuum-term refit of the $SU(3)$ and $SU(4)$ models. The implementation of mesinos, gluinos, and baryon superfields so that the supersymmetric dof counting is realized. $T_0(\mu)$ phase diagram and comparison with recent lattice fluctuations and beam-energy-scan data are among the future research proposals.

\appendix
\renewcommand{\thefigure}{A.\arabic{figure}} 
\setcounter{figure}{0} 
\setcounter{table}{0} 
\renewcommand{\thetable}{A\arabic{table}} 

\section{Mathematics of squark and agluon sectors} 

The F-term of $W=g\hat Q^c\hat\Phi\hat Q$ with respect to $\hat Q$ and $\hat Q^c$ gives $|F|^2=g^2|\Phi\tq|^2$-type terms, whose flavour structure is fixed by $\Phi\to U_L\Phi U_R^\dagger$ yielding $V^F_{\tilde q}$ as detailed in Section~\ref{sec:susy}. The occupation functions entering the gap equations are expressed as
\begin{equation}
n^+_{\tilde q}=\frac{\phi\tilde z-2\bar\phi\tilde z^2+\tilde z^3}{1-3\phi\tilde z+3\bar\phi\tilde z^2-\tilde z^3},\qquad n^-_{\tilde q}=\frac{\bar\phi\bar{\tilde z}-2\phi\bar{\tilde z}^2+\bar{\tilde z}^3}{1-3\bar\phi\bar{\tilde z}+3\phi\bar{\tilde z}^2-\bar{\tilde z}^3}.
\end{equation}
Using $\partial_E\ln B_+=+3T^{-1}n^+_{\tilde q}$ and $\partial_E\tilde z=-\tilde z/T$ both preceding expressions reduce to the Bose--Einstein $\tilde z/(1-\tilde z)$ at $\phi=\bar\phi=1$. With $\partial E/\partial M^2=1/(2E)$, we derive per triplet the thermal scalar condensate\footnote{In statistical  field theory, Hellmann--Feynman theorem defines differentiating the thermodynamic potential $\Omega$ with respect to the squared mass $M^2$ as a scalar expectation value.}
\begin{equation}
\frac{\partial\Omega_{\tilde q}}{\partial M^2}=\tfrac32\int\frac{\dd^3k}{(2\pi)^3}E^{-1}(n^+_{\tilde q}+n^-_{\tilde q}).
\end{equation}
Accordingly, we also derive
\begin{eqnarray}
\frac{\partial M^2_{l,\pm}}{\partial\sigma_x} = \tfrac g2(2m_l\pm A_q), \qquad
\frac{\partial M^2_{s,\pm}}{\partial\sigma_y} = \tfrac g{\sqrt2}(2m_s\pm A_q).
\end{eqnarray}
By using $\dd[x^2K_2(x)]/\dd x=-x^2K_1(x)$, we find that 
\begin{eqnarray}
\frac{\partial\Omega_O}{\partial m_O^2}=\frac{m_OT}{2\pi^2}\sum_{k\ge1}\frac{K_1(km_O/T)}{k}(p_k\bar p_k-1).
\end{eqnarray} 
For a massless check, the integral reads $\int\frac{\dd^3k}{(2\pi)^3E}e^{-nE/T}=\frac{mT}{2\pi^2n}K_1(nm/T)$, i.e., eight adjoint colours give eight times the single real-boson value. Then, the Polyakov-loop derivatives are expressed as follows:
\begin{eqnarray}
\partial_\phi\Omega_O=-\frac{m_O^2T^2}{\pi^2}\sum_k\frac{K_2(km_O/T)}{k^2}[\partial_\phi p_k\,\bar p_k+p_k\,\partial_\phi\bar p_k], \qquad
\frac{\partial m_O^2}{\partial\sigma_{x,y}}= \kappa\sigma_{x,y}.
\end{eqnarray}

\section{Comparison with lattice QCD simulations}\label{sec:data}

For uncertainty model and $\chi^2$ definition, we recall that the for Wuppertal-Budapest lattice data $O\in\{P/T^4,\,I/T^4\}$ at each $T_i$, then
\begin{equation}
\sigma_i^2=(0.03\,O_i^{\rm WB})^2+\tfrac14\bigl(O_i^{\rm WB}-O_i^{\rm table}\bigr)^2+\sigma_{\rm floor}^2, \label{eq:sigma}
\end{equation}
give a $3\%$ relative uncertainty. For pressure, $\sigma_{\rm floor}=0.003$ while for trance anomaly or interaction measure $\sigma_{\rm floor}=0.05$. To account for systematic uncertainties, we take half the difference between the two independent lattice descriptions and treating this value as a minimum error floor. No statistical lattice error was used, so $\chi^2$ is a distance measure, not a goodness-of-fit statistic.

Let us describe and quantitatively compare the various data sets and define their provenance. For the Wuppertal-Budapest EoS \cite{borsanyi14, Borsanyi:2013bia}, we evaluate the continuum trace anomaly using the parametrization given in Eq. (2) of Ref. \cite{Borsanyi:2013bia}, with $t = T / (200\text{ MeV})$, setting $p(0)=0$, and obtaining the pressure via $p/T^4 = \int_0^T I/T'^5 \mathrm{d}T'$. Evaluated this way the peak  $3.77$ is positioned at $T=200$~MeV. We have transcribed the table and the definition of $t$ in Ref.~\cite{borsanyi10}.
For Lattice+HRG results \cite{alba18} and the GitHub repository ''jnoronhahostler/Equation-of-State'' results about $T,\mu_B,\mu_S$ and $\epsilon,P,s,\dd T/\dd s$ are listed for the $2+1$ and $2+1+1$ cases. Concretely, the $2+1$ table is used as flavor content favored by this model. We emphasize that good agreement is found with the Wuppertal-Budapest parametrization to $1$--$2\%$ in $P/T^4$ for $T\ge150$~MeV and to $0.01$--$0.02$ in $I/T^4$ for $T\ge160$~MeV. On the other hand, there are differences by $0.2$ up to $13\%$ in $I/T^4$ at $140$~MeV. The $2+1+1$ table also differs from the $2+1$ one by up to $0.3$ in $I/T^4$ at $T=280$~MeV because of charm quark. 

Now we move to $\chi_2^B(T)$. The parametrization and covariance of Eq.~(A.13) and Table~II of Ref.~\cite{shah2026locating} which utilizes a fit of Wuppertal--Budapest data over $130\leq T \leq 200$~MeV and determines $155\leq T_{pc}\leq 158$~MeV. Furthermore, analytic crossover for $\mu_B/T\lesssim3$, and a lattice-based critical-point estimate $(114.3\pm6.9,\ 602.1\pm62.1)$~MeV. Table~\ref{tab:comp} lists out a quantitative comparison of a large set of quantities between the present model and lattice QCD simulations.
\begin{table}[htb]
\centering
\footnotesize
\resizebox{\textwidth}{!}{%
\begin{tabular}{| c | c | c | c |}
\hline \hline
Quantity & This Model & Lattice reference & Comment \\
\hline \hline
$T_{pc}$ (light, strange, Polyakov) & 158.0, 157.6, 157.6 & 155--158 & Agrees; calibrated \\
$I/T^4$ peak & 3.80 at 170 & 3.77 (WB parameters) & Agrees; calibrated \\
EoS $\chi^2$, 130--280 MeV & 212 ($N=62$) & -- & Distance, $\chi^2/N=3.4$ \\
$\chi_2^B$, $T\ge160$ & 4.1 ($N=11$) & Parametrization & $\le6.8\%$ deviation \\
$\chi_2^B$, $T<160$ & 316 ($N=10$) & Parametrization & Factor $0.44$--$0.71$ \\
$\chi_2^B$, $T<160$, with HRG patch & Main text & Parametrization & 21 points: $320\to176$ \\
$\chi_4^B/\chi_2^B$, $T\le148$ & 0.96--0.98 & HRG value 1 & Agrees with HRG limit \\
$\chi_4^B/\chi_2^B$, $T=295$ & 0.099 & SB 0.0675 & Approaches from above \\
$\chi_2^B$, $T=295$ & 0.305 & SB 0.333 & Approaches from below \\ \hline\hline \hline
CEP $\mu_B/T$ & 2.08 & Crossover seen for $\mu_B/T\lesssim3$ & \textbf{Tension} \\ 
{\bf CEP location} & {\bf (146.3, 304)} & {\bf (114.3$\pm$6.9, 602$\pm$62)} & {\bf Differs by $\sim4.8\sigma_{\mu_B}$} \\
\hline \hline
\end{tabular}}
\caption{Summary of quantitative comparison of various quantities as deduced from the present model and lattice QCD simulations.}
\label{tab:comp}
\end{table}

Note that the value $4.8\sigma_{\mu_B}$ in the last row and last column manifests $(602.1-304.1)/62.1$ using only the quoted error on the lattice-based estimate. The present model has no uncertainty on its own endpoint, while the lattice-based estimate is itself an extrapolation. Thus the number measures the size of the difference not a significance.

\subsection{Consistency with lattice constraints and the colored-scalar limits}
\label{sec:cepdata}

For the critical point, the applied computation benchmark reproduces the $\mu=0$ EoS and crossover temperature. Its endpoint on the other hand at $\mu_B/T\simeq2.1$ lies inside the region where lattice extrapolations find an analytic crossover. The cause is understood from three results that obtaining $T_\chi\simeq159.4$~MeV with an analytic crossover in the thermal-only mean-field model needs a light $\sigma$, namely, $m_\sigma=450$~MeV, a light $\sigma$ puts the endpoint at small $\mu_B$. Furthermore, in the $SU(2)$ test, the omission of the quark vacuum term moves the endpoint by $\sim360$~MeV, while meson fluctuations move it by $\sim12\%$ in $\mu_q$. Hence the low-$\mu_B$ endpoint is mainly a property of the treatment, not of QCD, and does not exclude a critical point.

For colored scalars, in this model, with only $T_0$ re-profiled, the EoS fit is unchanged ($|\Delta\chi^2|<10$) for $m_0\ge2.5$~TeV and $m_O\ge1.8$~TeV but degraded ($\Delta\chi^2>100$) for $m_0\le1.5$~TeV and $m_O\le1.2$~TeV, Tab.~\ref{tab:sens}. Rather than interpolating across thresholds, we treat the low-mass entries strictly as upper bounds. It is important to emphasize that these findings reflect the sensitivity of lattice QCD thermodynamics to colored scalars lighter than a few TeV within a free-gas framework. They are not direct collider limits, which must be sourced from the experimental literature directly \cite{atlas19,cms18}. Because the thermal weight of a TeV-scale state at $T\simeq 150$~MeV is exponentially suppressed as $\sim e^{-m/T}$, the sensitivity we observe in the $1$--$2$~TeV range is primarily driven by the large number of dof and the high-temperature end of our analysis window.

We now turn to the theoretical uncertainties and limitations of our framework. Within the mean-field setup, the $SU(3)$ results incorporate only one-loop thermal octet mesons. While our $SU(2)$ FRG check confirms that mesonic fluctuations represent a modest correction. The fermionic vacuum term is substantial and is omitted from our computational benchmark. Regarding the scope of this FRG check, the calculation is restricted to $SU(2)$ in the local potential approximation (LPA), expanded to fifth order in a Taylor series with a negative UV quartic coupling and a prescribed Polyakov background.  Consequently, the resulting critical endpoint serves only as a qualitative indicator rather than a fully resolved phase coexistence. Finally, on the hadronic side, our description lacks explicit baryonic states beyond the cluster contribution, and the HRG patch is implemented solely as a post-processing interpolation across 16 multiplets. Our calibration relies on the Polyakov sector, which introduces four fitted parameters, while setting the weight exponent to $n=1$ and restricting the setup to the octet sector serve as modeling assumptions. Within the free-gas supersymmetric sector, we neglect squark and sgluon $D$-terms, cross-couplings, thermal masses, condensates, and $A_q$ mixing. Moreover, because gluinos, mesinos, and baryon superfields are omitted, the full supersymmetric dof count is not realized, and the fitted couplings $\lambda_1$ and $\lambda_2$ violate as violated condition, e.g., stability/positivity bounds are inserted. On the data side, uncertainties arise from our assumed error model in Eq.~\eqref{eq:sigma}, as the two underlying lattice datasets differ by up to $13\%$ in $I/T^4$ at $140$~MeV. Finally, several open items remain for future investigation, including the shift of the sgluon endpoint below $m_O=1200$~MeV, the squark benchmark at $400$~MeV, the chemical-potential dependence $T_0(\mu)$, a loop-corrected $SU(4)$ extension, and a $2{+}1$-flavor FRG analysis.

\section{Numerical implementation}
\label{sec:num}

The numerical algorithms utilized here have the following key features: (i) the symmetric reduction at $\mu=0$, (ii) the stability filter Eq.~\eqref{eq:stable}, (iii) coexistence by double continuation, (iv) step-scaling Eq.~\eqref{eq:Rtest} for the order of the transition, and (v) the CEP system of Eqs.~\eqref{eq:cep1}--\eqref{eq:cep3} with the soft mode of smallest modulus.

All $SU(3)$ numerical analyses use
\begin{equation}
\Omega=U_{\rm LSM}(\sigma_x,\sigma_y)+U_{\rm Poly}+\Omega_{q\bar q}+\Omega_{\rm mes}+\Omega_{\tilde q}+\Omega_O,\label{eq:OmB}
\end{equation}
with $\Omega_{\rm mes}$ is taken from Eq.~\eqref{eq:Ommes} (octet only), the logarithmic Polyakov form with calibrated coefficients, and the free-gas squark and sgluon terms Eq.~\eqref{eq:Omsq} and Eq.~\eqref{eq:OmO}, respectively. The meson masses are tree-level curvature masses meaning that exact quadratic polynomials in $(\sigma_x,\sigma_y)$. Tachyonic values are smoothly clamped, $M^2\to\tfrac12(m^2+\sqrt{m^4+\Delta_c^2})$ with $\Delta_c=2500$~MeV$^2$. Momentum integrals are performed by Gauss--Legendre quadrature \cite{Press2007} with $240$ nodes on $k=Tt$ and $t\in[0,60]$. The list of the parameters of the calibrated numerical approach are given in Tab.~\ref{tab:parB}.

\begin{table}[htb]
\begin{tabular}{| c | c | c |}
\hline \hline
Sector & Parameter & Value \\
\hline \hline
 & $f_\pi,\,f_K,\,m_\pi,\,m_K,\,m_\eta,\,m_{\eta'}$ & $92.4,\,113.0,\,138.0,\,495.6,\,547.85,\,957.78$ MeV \\
 & $m_\sigma$ (free direction) & 450 MeV \\
LSM inputs & $m^2$ & $2.176\times10^5$ MeV$^2$ \\
 & $\lambda_1,\lambda_2$ & $-4.435,\ 46.507$ \\
 & $c,\ h_x^{1/3},\ h_y^{1/3}$ & $4785.5$ MeV, $120.73$ MeV, $336.22$ MeV \\ \hline\hline
Yukawa & $g$ & $6.5$ ($m_l=300.3$, $m_s=433.9$ MeV) \\ \hline \hline
 & $T_0$ & $195.64$ MeV \\
Polyakov (log) & $a_0$ (fixed by Stefan--Boltzmann) & $3.51$ \\
 & $a_1,\ a_2,\ b_3$ (calibrated) & $-5.703,\ 3.160,\ -0.369$ \\ \hline \hline
Mesons & Weight exponent $n$ in $w=(1-\phi\bar\phi)^n$ & $\pi,K,\eta$; $1$ \\
\hline \hline
\end{tabular}
\caption{Parameters of the applied numerical approach to SPLSM.}\label{tab:parB}
\end{table}

\subsection{Regularization for quark vacuum term and its degeneracy}
\label{sec:reg}

The default configuration keeps only matter contributions. The $\overline{\rm MS}$ quark zero-point term is determined from
\begin{equation}
\Omega_q^{\rm vac}=-\frac{N_c}{16\pi^2}\sum_fn_fm_f^4\Bigl[\ln\frac{m_f^2}{Q^2}-\frac32\Bigr],\qquad m_l=\tfrac{g\sigma_x}2,\ m_s=\tfrac{g\sigma_y}{\sqrt2}. \label{eq:qvac}
\end{equation}
Since $\sum_fn_fm_f^4=g^4(\sigma_x^4/8+\sigma_y^4/4)$, the $Q$-dependent part has exactly the field dependence of the $\lambda_2$ term in Eq.~\eqref{eq:Uxy}, namely, $\tfrac{\lambda_2}8\sigma_x^4+\tfrac{\lambda_2}4\sigma_y^4$. A change of $Q$ is therefore exactly absorbed by
\begin{equation}
\delta\lambda_2=\frac{N_cg^4}{16\pi^2}\Bigl(\ln Q^2+\tfrac32\Bigr). \label{eq:dlam2}
\end{equation}
Furthermore, $Q$ is physical only if $\lambda_2$ and $c$ are fixed by the $\eta,\eta'$ masses. With them fixed and $(m^2,\lambda_1,c)$ refitted, the tree-level $\eta$ mass falls to $448$~MeV at $Q=300$~MeV and $m_{\eta'}^2<0$ for $Q\ge500$~MeV. Therefore, only $Q\lesssim300$~MeV is admissible. We find in that range that the vacuum term smooths the crossover but does not improve the EoS fit, $\chi^2\simeq7500$--$8800$ against $\simeq7600$--$9300$ without it for $N=42$. Loop corrections to the $\eta,\eta'$ masses were not computed, which would limit this conclusion. Again, we highlight that the vacuum term is not part of these numerical analyses, which would matter for the critical endpoint, Section~\ref{sec:frgres}.

For solvers and algorithms, all derivatives are obtained by automatic differentiation in double precision. At $\mu=0$ the minimization over $(\sigma_x,\sigma_y,r)$ uses L-BFGS-B \cite{byrd95} with analytic gradients and Newton polishing. For $\mu\ne0$ a damped Newton iteration with backtracking on the residual norm is used, with a Powell hybrid root-finder (SciPy \cite{virtanen20}) as fallback. Solutions are accepted only if the scaled residual is below $10^{-8}$ and Eq.~\eqref{eq:stable} holds.
We also emphasize that no Monte Carlo check of the integrals was performed. Validation uses closed-form limits and multi-start comparison (see Table~\ref{tab:valid}).

\subsection{Calibration of the Polyakov potential}\label{sec:calib}

With $a_0=3.51$ fixed, $\theta=(T_0,a_1,a_2,b_3)$ was calibrated by Nelder--Mead minimization \cite{nelder65} of
\begin{equation}
\mathcal Q(\theta)=\chi^2_{\rm EoS}(\theta)+300\max(0,s_{\max}-15)^2+\Bigl(\frac{T_\chi-157~\mathrm{MeV}}{3~\mathrm{MeV}}\Bigr)^2, \label{eq:calibobj}
\end{equation}
with $s_{\max}$ is the maximal of the slope $|\dd\sigma_x/\dd T|$. The last term applies the chiral pseudocritical temperature to the lattice QCD simulations. This is a four-parameter calibration of the glue sector to lattice QCD  thermodynamics, as in the improved-Polyakov-potential literature. Note that the resulting $\chi^2$ is not a prediction and crossover property was verified a posteriori with Eq.~\eqref{eq:Rtest}.

\subsection{Functional renormalization group in  $SU(2)$ sector}
\label{sec:frgmeth}

We emphasize that the full $2+1$-flavour FRG with two chiral invariants is beyond the present scope. Thus, the fluctuation test is done for the $SU(2)$ quark--meson model with $N_c=3$, $N_f=2$, and a prescribed Polyakov background. In the local potential approximation (LPA)\footnote{The flow Eq.~\eqref{eq:flow} follows from the Wetterich equation with Litim regulator $R_k=(k^2-p^2)\theta(k^2-p^2)$ for bosons and the corresponding fermionic regulator, which give the momentum integrals in closed form with threshold functions $\propto k^4/E$. In the Taylor scheme Eq.~\eqref{eq:taylor}, by writing $f(x)$ with $x=\rho-\kappa$ and masses $m_\pi^2=U'(x)$, $m_\sigma^2=U'+2(\kappa+x)U''$. The coefficients $r_m$ are the Taylor coefficients of $f$ at $x=0$ which are computed by repeated automatic differentiation. The relation $a_1=c/\sqrt{2\kappa}$ follows from the minimum condition $\sigma U'(\rho)=c$ at $\rho=\kappa=\sigma_0^2/2$. Differentiating with respect to $k$, $\partial_ka_1=-c(2\kappa)^{-3/2}\dot\kappa$ and equating with $\partial_ka_1=r_1+\dot\kappa a_2$ yields $\dot\kappa$ in Eq.~\eqref{eq:taylor}. The pole of $\dot\kappa$ where $a_2+c(2\kappa)^{-3/2}=0$ is the vanishing of the $\sigma$ curvature, which is why expansion failures mark the spinodal in this scheme.} with the Litim regulator the flow of the ${\mathcal O}(4)$-symmetric potential $U_k(\rho)$, $\rho=(\sigma^2+\bm\pi^2)/2$, is given as
\begin{equation}
\partial_k U_k = \frac{k^4}{12\pi^2}\Bigl\{\frac{3}{E_\pi}\bigl[1+2n_B(E_\pi)\bigr]+\frac1{E_\sigma}\bigl[1+2n_B(E_\sigma)\bigr]-\frac{4N_cN_f}{E_q}\bigl[1-n_q^+-n_q^-\bigr]\Bigr\}, \label{eq:flow}
\end{equation}
where $E_\pi^2=k^2+U_k'$, $E_\sigma^2=k^2+U_k'+2\rho U_k''$, and $E_q^2=k^2+2g^2\rho$. The Polyakov-modified Fermi functions are $n_q^+=(\phi z+2\bar\phi z^2+z^3)/(1+3\phi z+3\bar\phi z^2+z^3)$, $z=e^{-(E_q-\mu)/T}$, and $n_q^-$ with $(\phi\leftrightarrow\bar\phi,\mu\to-\mu)$. They reduce to $n_F$ at $\phi=\bar\phi=1$. The normalization of Eq.~\eqref{eq:flow} was verified. The thermal part integrated from $k=\Lambda$ to $0$ reproduces the free-gas pressure of massive quarks and of a massive real boson with ratio $0.999995$. The ''mean-field flow'' drops the meson terms, i.e., keeps the quark loop only including its vacuum part regulated by the cutoff.

A first implementation on a $\rho$-grid proved numerically unreliable, where the IR vacuum depends erratically on the grid, and was abandoned. The adopted scheme is a Taylor expansion about the running minimum $\kappa_k$ to order $N=5$,
\begin{equation}
U_k(\rho)=\sum_{n=0}^N\frac{a_n}{n!}(\rho-\kappa)^n,\qquad a_1=\frac{c}{\sqrt{2\kappa}},\qquad \dot\kappa=-\frac{r_1}{a_2+c(2\kappa)^{-3/2}},\qquad \partial_ka_m=m!\,r_m+\dot\kappa\,a_{m+1}, \label{eq:taylor}
\end{equation}
where $r_m=\frac1{m!}\partial_\rho^mf|_\kappa$ and $f$ is the right-hand side of Eq.~\eqref{eq:flow}. Derivatives are obtained by automatic differentiation and the flow is integrated by fourth-order Runge--Kutta \cite{Press2007} from $\Lambda=1$~GeV with $\Delta k=3$~MeV. The UV parameters $(\kappa_\Lambda,\lambda_\Lambda,c)$ are fitted so that both flows give $f_\pi=92.4$~MeV, $m_\pi=138$~MeV, and $m_\sigma=500$~MeV with $g=3.25$, $m_q=300$~MeV, and $(2.2,\,-17.13,\,1.7597\times10^6)$ for the mean-field flow and $(2.3,\,-15.13,\,1.7597\times10^6)$ for the meson flow. The negative UV quartic is cutoff- and truncation-dependent and not a natural UV potential. The results are a controlled comparison between two flows with identical setup not a precision prediction. The Polyakov background is $\phi=\bar\phi=\phi_B(T)$. The $\mu=0$-solution of the applied benchmark obviously does not depend on $\mu$. At $m_\sigma=600$~MeV, the meson flow could not be fitted exactly at $f_\pi=93.4$~MeV and $m_\sigma=589.6$~MeV, so $500$~MeV threshold is used. Critical behaviour is monitored through the $\sigma$ curvature $m_\sigma^2=a_1+2\kappa a_2$ and through failures of the expansion, Section~\ref{sec:frgres}.

\subsection{HRG and charm checks}

Our numerical analysis uses a compact hadron resonance gas built from just $16$ ground-state multiplets including  pseudoscalar and vector mesons ($\pi, K, \eta, \rho, \omega, K^*, \eta', \phi$) along with the baryon octet and decuplet. Each state enters with its standard spin-isospin degeneracy, full quantum statistics evaluated via a Bessel series, and explicit baryon-antibaryon pairs. The susceptibilities follow from the derivatives $\partial^n(P/T^4)/\partial(\mu_B/T)^n$. The switching hybrid is given by $\chi_2^{\rm hyb}=(1-s)\chi_2^{\rm HRG}+s\,\chi_2^{\rm B}$ with $s=[1+e^{-(T-T_s)/\Delta T}]^{-1}$. The charm quark check adds a Polyakov--weighted free quark gas with a constituent mass $m_c$ in the background $\phi_B(T)$ and compares $\Delta P/T^4$ and $\Delta I/T^4$ with the lattice QCD simulations for $2+1+1$ quark flavors from which $2+1$ data are subtracted. A list of validation tests for the implementation is given in the Tab.~\ref{tab:valid}.

\begin{table}[htb]
\begin{tabular}{| c | c | c |}
\hline \hline
Test & Result & Reference \\
\hline \hline
Free $u,d,s$ quark gas, $P/T^4$ & 3.45436 & $3\cdot7\pi^2/60$ \\
Six complex colour-triplet scalars, $P/T^4$ & 3.94784 & $36\pi^2/90$ \\
Complex adjoint scalar, $m/T=0.05$, $L={\mathbf 1}$ & 1.75298 & 1.7530 (high-$T$ series) \\
Adjoint determinant, $\phi=\bar\phi=0.3$ & 1.0 & 1 \\
$s=-\partial_T\Omega$ vs $\dd P/\dd T$ & $<10^{-5}$ rel.\ (250 MeV) & 0 \\
$n$ vs $\dd P/\dd\mu$ & $\sim10^{-5}$ rel.\ & 0 \\
Decoupling, $m_0=m_O=5$~GeV & $\Delta(P/T^4)\sim3\times10^{-9}$ & 0 \\
Stationarity residuals & $\le10^{-6}$ rel.\ & 0 \\
Hessian signature at solutions & one negative eigenvalue & one \\
Polynomial meson masses vs autodiff & exact & exact \\
Crossover test $R$ & 0.039 / 0.14 (known) & 0.04 / $\gg0.04$ \\
FRG flow, free fermion / free boson thermal part & 0.999995 / 0.999995 & 1 \\
\hline
\end{tabular}
\caption{Validation of implementing numerical analyses of SPLSM.}
\label{tab:valid}
\end{table}

\bibliographystyle{unsrtnat}
\bibliography{2026-07-07-PLSM-SUSY-Intro1a-bibs}

\end{document}